\documentclass[showpacs,preprintnumbers,amsmath,amssymb,nofootinbib]{revtex4}
\usepackage{graphicx}% Include figure files
\usepackage{dcolumn}% Align table columns on decimal point
\usepackage{bm}% bold math
\usepackage{slashed}
\usepackage{multirow}
\usepackage{color}
\usepackage{ulem}

\begin{document}

\title{Revisiting the $b_1\pi$ and $\rho\pi$ decay modes of the $1^{-+}$ light hybrid state with light-cone QCD sum rules}

\author{Zhuo-Ran Huang$^{1,2}$, Hong-Ying Jin$^1$, T.G. Steele$^2$ and Zhu-Feng Zhang$^3$\\
$^1$Zhejiang Institute of Modern Physics, Zhejiang University, Zhejiang Province, 310027, P. R. China\\
$^2$Department of Physics and Engineering Physics, University of Saskatchewan, Saskatoon, Saskatchewan, S7N 5E2, Canada\\
$^3$Physics Department, Ningbo University, Zhejiang Province, 315211, P. R. China
}

\begin{abstract}
We study the $\rho\pi$ and $b_1\pi$ decay modes of the $1^{-+}$ light hybrid state within the framework of light-cone QCD sum rules.
We use both the tensor current $\bar\psi\sigma_{\mu\nu}\psi$ and the derivative current $\bar{\psi}\overleftrightarrow{D}_\mu\gamma_5\psi$ as interpolating currents to calculate the partial decay width of the $b_1\pi$ decay mode. Comparing the sum rules obtained by using different currents, we obtain  $\Gamma(\pi_1\to b_1\pi)$ = 8--23, 32--86 and 52--151\,MeV for $m_{1^{-+}}$ = 1.6, 1.8 and 2.0\,GeV respectively, which favour the results from the flux tube models and lattice simulations. We also use the tensor current to study the $\rho\pi$ decay mode, and although an extended stability criterion is needed, our results suggest a small partial decay width.
\end{abstract}

%12.38.Lg   Other nonperturbative calculations
%12.39.Mk   Glueball and nonstandard multi-quark/gluon states
%14.40.Rt   Exotic mesons

\maketitle

\section{Introduction}

The $1^{-+}$ light hybrid mesons have attracted particular attention in hadronic physics. The reason is not only such a state can be distinguished from ordinary $q\bar q$ mesons for its beyond-quark-model exotic quantum number but also it is expected to be one of the lowest-lying hybrid states. To date, accumulated experimental data have shown the existence of $1^{-+}$ isovector states, i.e. $\pi_{1}(1400)$, identified in $\eta\pi$ and $\eta'\pi$ channels and $\pi_{1}(1600)$ seen to decay into $b_1\pi$, $f_1\pi$ and $\eta'\pi$ \footnote{The situation is uncertain for the $\rho\pi$ decay mode: VES and Compass have not claimed the existence of the $\rho\pi$ decay while some people have argued that the phase motion results observed by E852 can be resulted from the leakage of  $\pi_2(1670)$ \cite{Agashe:2014kda,Meyer:2015eta}.} \cite{Agashe:2014kda}.
Moreover, there is another $1^{-+}$ state, $\pi_1(2015)$, quoted in the extended version of PDG \cite{Agashe:2014kda}, which is only observed by E852 in $f_1\pi$ and $b_1\pi$ final states and needs further confirmation.

Computations on the light hybrid spectrum have been conducted with lattice QCD and different phenomenological models (for a review, see \cite{Meyer:2015eta}). In the bag model, the predicted mass of the low lying $1^{-+}$ hybrid nonet is around 1.5\,GeV \cite{Chanowitz:1982qj,Barnes:1982tx}. The earliest quenched lattice calculations predicted the $1^{-+}$ light hybrid mass lies in the region 1.8--2.1\,GeV \cite{Lacock:1998be,McNeile:1998cp,Mei:2002ip,Hedditch:2005zf}, while the more recent dynamical calculations predicted the mass is around 2.2\,GeV \cite{Dudek:2010wm,McNeile:2006bz}. Isgur and Paton estimated in the flux tube model the $1^{-+}$ light hybrid mass to be 1.9\,GeV \cite{Isgur:1983wj,Isgur:1984bm} while in the constituent gluon models, the exotic light hybrid mass are found to lie in the region 1.8--2.2\,GeV \cite{LeYaouanc:1984gh,Ishida:1991mx,Iddir:1988jd}.

In the framework of QCD Sum Rules \cite{Shifman:1978bx}, the earliest leading-order results obtained by different authors show the $1^{-+}$ light hybrid mass lies in the range 1.6--2.1\,GeV \cite{key-6-1,key-7-1,key-8-1,key-8-2,key-8-3,key-8-4,key-8-5}. Over the past 15 years, different groups extended and improved the sum rule calculation. The radiative corrections were calculated in \cite{key-8-6} and \cite{key-9} for the perturbative terms and in \cite{key-9-1} for the OPE. The short distance tachyonic gluon mass effects were also included in \cite{key-8-6}, and Narison gave a systematical re-examination of the $1^{-+}$ mass with inclusion of all the previous calculated effects and estimated the mass to be 1.81\,GeV \cite{Narison:2009vj}. Furthermore, the authors of this paper further complemented the sum rule analysis of the $1^{-+}$ mass: instanton effects were studied in Zhang's PhD thesis \cite{zhangthesis}, and a monte-carlo based uncertainty analysis was performed in \cite{Zhang:2013rya}. Both of these efforts show little change in the mass prediction. Moreover, recently we included the higher power corrections of OPE with consideration of operator renormalization \cite{Huang:2014hya}. We considered  violation of factorization of higher dimensional condensates and updated the QCD input parameters. We obtained a quite conservative range of the $1^{-+}$ light hybrid mass, i.e. 1.72--2.60\,GeV, which only covers $\pi_1(2015)$ and does not support $\pi_1(1600)$ as a pure hybrid. Given that the analysis in \cite{Huang:2014hya} has involved all effects that seem to have considerable influence in the sum rule mass extraction, the mass range can be considered as a general conclusion from QCD sum rules.

From the theoretical mass predictions we can see that $\pi_1(1400)$ is not supported to be a hybrid by various theoretical schemes. Even the mass of $\pi_1(1600)$ is lower than many of the theoretical predictions, although this resonance has long been considered as a good hybrid candidate. Some people have argued that $\pi_1(1600)$ can involve a four-quark state, and a mixing of molecular state and four-quark state has been proposed in \cite{Narison:2009vj} based on the discussions in \cite{Chen:2008qw} and \cite{Zhang:2004nb} about $1^{-+}$ tetraquark and molecular states. The unconfirmed $\pi_1(2015)$ has been suggested in \cite{Narison:2009vj} and \cite{Huang:2014hya} to be a good hybrid candidate. To shed further light on the nature of these states needs both theoretical and experimental study of the $1^{-+}$ decay modes.

The UKQCD collaboration examined the decay of $1^{-+}$ hybrid with two dynamical quarks in the lattice simulation, and obtained the partial decay widths $\Gamma(\pi_1\rightarrow b_1\pi)=400\pm120$\,MeV and $\Gamma(\pi_1\rightarrow f_1\pi)=90\pm60$\,MeV \cite{McNeile:2006bz}. Later in \cite{Burns:2006wz}, Burns and Close found these results agree quite well with the predictions near threshold in the flux tube model, thus they reduced the partial widths to $\Gamma(\pi_1\rightarrow b_1\pi)\approx80$\,MeV and $\Gamma(\pi_1\rightarrow f_1\pi)\approx25$\,MeV, whereas the results in IKP Model \cite{Kokoski:1985is,Isgur:1985vy} and PSS Model \cite{Page:1998gz,Swanson:1997wy} are $\Gamma(\pi_1\rightarrow b_1\pi)=51$\,MeV, $\Gamma(\pi_1\rightarrow f_1\pi)=14$\,MeV and $\Gamma(\pi_1\rightarrow b_1\pi)=40-78$\,MeV, $\Gamma(\pi_1\rightarrow f_1\pi)=10-18$\,MeV respectively. The decay modes of the $1^{-+}$ light hybrid state have also been studied within the framework of QCD sum rules. The earliest three-point function sum rule studies can be seen in \cite{DeViron:1985xn} and \cite{key-8-3}, while a recent study can be seen in \cite{Chen:2010ic} where the pion mass terms in the denominator were ignored and only the $1/q^2$ terms divergent in the limit $q^2\rightarrow0$ were kept. The authors in \cite{Chen:2010ic} also studied the $1^{-+}$ decay \cite{Huang:2010dc} using the light-cone QCD sum rules (LCSR) \cite{Balitsky:1989ry,Braun:1988qv,Chernyak:1990ag,Belyaev:1994zk}, of which the basic idea is to expand the correlation function near the light cone. However, the predicted partial decay width of $\pi_1\rightarrow b_1\pi$ is somewhat confusing. The suggestion of very tiny decay width of $b_1\pi$ implies both $\pi_1(1600)$ and $\pi_1(2015)$ may not have much of a hybrid constituent, and also doesn't agree with the predictions from various models mentioned above.
Considering the important role of $b_1\pi$ decay mode in identifying the $1^{-+}$ hybrid state, it's worthwhile to re-examine this decay mode within the same theoretical framework.

In this work, we study the $b_1\pi$ decay mode using the $b_1$ derivative current $\bar{\psi}\overleftrightarrow{D}_\mu\gamma_5\psi$ instead of the current $\bar{\psi}\overleftrightarrow{\partial}_\mu\gamma_5\psi$ adopted in \cite{Huang:2010dc}. We also use the tensor current $\bar\psi\sigma_{\mu\nu}\psi$, which not only couples to $b_1$ but also the $1^{--}$ $\rho$ meson, thus an analysis of the $\rho\pi$ decay mode can be provided simultaneously. Usually, the $\rho$ meson is studied using the simpler vector interpolating current  $\bar \psi\gamma_\mu \psi$, as was done in \cite{Shifman:1978bx,Leupold:1997dg} for the mass and in \cite{AliKhan:2001xoi,Jansen:2009hr} for the decay constant. In addition, attempts to study the $\rho$ meson using the tensor current have also been made previously in both sum rule \cite{Bakulev:1999gf,reinder} and lattice calculations \cite{Jansen:2009hr}. Studies using the tensor current can provide a useful re-examine of the results obtained by using the vector current. Previous sum rule studies using the vector current $\bar \psi\gamma_\mu \psi$ predict large partial decay width of $\rho\pi$ channel \cite{Chen:2010ic,Huang:2010dc} while this channel is forbidden in the original flux tube model \cite{Isgur:1985vy} and the partial decay width is still small in its modified versions \cite{Page:1998gz,Swanson:1997wy,Close:1994hc}, therefore, it is also worth re-examining this channel by using the tensor current.

We arrange the article as follows: In Sec.~\ref{LCQSR} we illustrate the formalism of the light-cone QCD sum rules for deriving the coupling constants in the $1^{-+}$ decay amplitudes. In Sec.~\ref{LCexpansion} we present our results of the light-cone expansion of the correlation function of both the tensor current $\bar\psi\sigma_{\mu\nu}\psi$ and the derivative current $\bar{\psi}\overleftrightarrow{D}_\mu\gamma_5\psi$. In Sec. \ref{integral} we illustrate the method of calculating the integrals of the spectral densities, from which the contribution from excited states and continuum can be subtracted. In Sec.~\ref{numerical}, we present the numerical analysis of $b_1\pi$ and $\rho\pi$ decay modes with both currents. In Sec.~\ref{summary} we present the summary and conclusions.

\section{light-cone QCD sum rules for the $1^{-+}$ light hybrid state}
\label{LCQSR}

We begin with the following correlation function to study the decay modes $\pi_1\rightarrow b_1\pi$ and $\pi_1\rightarrow \rho\pi$:
\begin{equation}
\label{eq:master1}
\Pi^{T,D}(k,p)=i\int d^4 xe^{ik\cdot x}\langle \pi(q)|T\{J^{T,D}(x)J^{H^\dagger}(0)\}|0\rangle,
\end{equation}
where $p$, $k$ and $q$ are respectively the momentum for $\pi_1$, $b_1$ or $\rho$ and $\pi$, which satisfy the four-momentum conservation $p=k+q$.\ $J^H=J^H_\mu=\bar{\psi}G_{\mu\nu}\gamma_{\nu}\psi$ couples to the $1^{-+}$ light hybrid, $J^T=J^T_{\mu\nu}=\bar\psi\sigma_{\mu\nu}\psi$ couples to $b_1$ and $\rho$, and $J^D=J^D_\mu=\bar{\psi}\overleftrightarrow{D}_\mu\gamma_5\psi$ also couples to $b_1$.

In the practical calculation, we use $J^{H}_\mu=\frac{\sqrt2}{2}(\bar{u}G_{\mu\nu}\gamma_{\nu}u-\bar{d}G_{\mu\nu}\gamma_{\nu}d)$, $J^T_{\mu\nu}=\bar{d}\sigma_{\mu\nu}u$ and $J^D_\mu=\bar{d}\overleftrightarrow{D}_\mu\gamma_5u$ to study the partial decay widths of decay modes $\pi_1^0\rightarrow b_1^+\pi^-$ and $\pi_1^0\rightarrow \rho^+\pi^-$, of which the results also hold for $\pi_1^0\rightarrow b_1^-\pi^+$ and $\pi_1^0\rightarrow \rho^-\pi^+$. We define the decay constants through the following formulas:
\begin{eqnarray}
\label{eq:2}
\langle 0|J^{H}_\mu(0)|\pi_1\rangle=f_{\pi_1}m_{\pi_1}^3\eta_\mu\,,\langle0|J^T_{\mu\nu}(0)|b_1\rangle=if^T_{b_1}\varepsilon_{\mu\nu\rho\sigma}\epsilon^\rho k^\sigma,\\\nonumber
\langle0|J^T_{\mu\nu}(0)|\rho\rangle=if^T_{\rho}(k_\mu\epsilon_\nu-k_\nu\epsilon_\mu)\,,\langle0|J^D_\mu(0)|b_1\rangle=f_{b_1}\epsilon_\mu,
\end{eqnarray}
where $\epsilon_\mu$ and $\eta_\mu$ are polarization vectors, and the decay amplitudes can be written as:
\begin{eqnarray}
\label{eq:3}
\mathcal{M}(\pi_1\rightarrow\rho\pi)
=ig_\rho\varepsilon_{\alpha\beta\rho\sigma}\epsilon^{*\alpha}\eta^\beta k^\rho p^\sigma,\\\nonumber
\mathcal {M}(\pi_1\rightarrow b_1\pi)
=ig_{b_1}^1(\eta\cdot\epsilon^*)+ig_{b_1}^2(\eta\cdot k)(\epsilon^*\cdot p).
\end{eqnarray}

The correlation function can be expanded in the light-cone distribution amplitudes which play the similar role as the condensates of local operators in the SVZ operator product expansion. The light-cone expansions can be compared to the phenomenological expressions of the correlation function so as to estimate the coupling constants in \eqref{eq:3} and then to obtain the partial decay widths. After interpolating the intermediate hadronic states into \eqref{eq:master1} and using the definitions in \eqref{eq:2} and \eqref{eq:3}, we arrive at the phenomenological sides:
\begin{eqnarray}
\label{eq:4}
\Pi^T_{b_1}(k,p)&=&i\int d^4 xe^{ik\cdot x}\langle \pi^-(q)|T\{J^{T}_{\mu\nu}(x)J^{H^\dagger}_\alpha(0)\}|0\rangle\nonumber\\
&\rightarrow&{f_{\pi_1} f^T_{b_1}\over (p^2-m_{\pi_1}^2) (k^2-m_{b_1}^2)}\epsilon_{\mu\nu\rho\sigma}\biggl[g^1_{b_1}k^\rho(-\frac{p_\alpha p^\sigma}{p^2}+g_\alpha^{\ \sigma})+g^2_{b_1}k^\rho(-\frac{k\cdot p}{p^2}p_\alpha p^\sigma+k_\alpha p^\sigma)\biggl]+\cdots,
\end{eqnarray}
\begin{eqnarray}
\label{eq:5}
\Pi^T_{\rho}(k,p)&=&i\int d^4 xe^{ik\cdot x}\langle \pi^-(q)|T\{J^{T}_{\mu\nu}(x)J^{H^\dagger}_\alpha(0)\}|0\rangle\nonumber\\
&\rightarrow&g_\rho{f_{\pi_1} f^T_{\rho}\over (p^2-m_{\pi_1}^2) (k^2-m_{\rho}^2)}(-\epsilon_{\rho\sigma\nu\alpha}k^\rho p^\sigma k_\mu+\epsilon_{\rho\sigma\mu\alpha}k^\rho p^\sigma k_\nu)+\cdots,
\end{eqnarray}
\begin{eqnarray}
\label{eq:6}
\Pi^D_{b_1}(k,p)&=&i\int d^4 xe^{ik\cdot x}\langle \pi^-(q)|T\{J^{D}_\mu(x)J^{H^\dagger}_\nu(0)\}|0\rangle\nonumber\\
&\rightarrow&{if_{\pi_1} f_{b_1}\over (p^2-m_{\pi_1}^2) (k^2-m_{b_1}^2)}\biggl[g^1_{b_1}(\frac{k\cdot pk_\mu p_\nu}{k^2p^2}-\frac{k_\mu k_\nu}{k^2}-\frac{p_\mu p_\nu}{p^2}+g_{\mu\nu})\\\nonumber
&&+g^2_{b_1}(\frac{{(k\cdot p)}^2k_\mu p_\nu}{k^2p^2}-\frac{k\cdot p}{k^2}k_\mu k_\nu-\frac{k\cdot p}{p^2}p_\mu p_\nu+p_\mu k_\nu)\biggl]+\cdots,
\end{eqnarray}
where the ellipses denote the contribution from excited states and continuum.

On the QCD side, the correlation function \eqref{eq:master1} can be expanded near the light-cone $x^2 = 0$ in terms of meson distribution amplitudes of different twists. After picking out characteristic tensor structures we get invariant parts of correlation functions corresponding to different coupling constants in \eqref{eq:4}--\eqref{eq:6}. Sometimes this process involves some technical complications as different tensor structures entangle with each other at first sight. We will discuss these details in the next section.

In order to subtract the contribution from excited states and continuum in the invariants of correlation functions, the double dispersion relation can be used:
\begin{eqnarray}\label{dispersionrelation}
\label{eq:7}
\Pi(k^2,p^2)
=\int_0^\infty ds_1\int_0^\infty ds_2 \frac{\rho(s_1,s_2)}{(s_1-k^2-i\epsilon)(s_2-p^2-i\epsilon)}
+\textrm{subtractions},
\end{eqnarray}
where the subtractions eliminate the infinities from the dispersion integral. After taking Borel transformation,
which is defined as
\begin{eqnarray}
\label{eq:10}
\mathcal{B}_{k^2}^{M^2}[f(k^2)]
=\lim_{n\rightarrow\infty}\frac{(-k^2)^{n+1}}{n!}\left(\frac{d}{dk^2}\right)^nf(k^2)\left|_{k^2=-nM^2}\right.,
\end{eqnarray}
the subtraction terms can be removed and then we get
\begin{eqnarray}
\label{eq:8}
\mathcal{B}_{k^2}^{\frac{1}{\sigma_1}}\mathcal{B}_{p^2}^{\frac{1}{\sigma_2}}\Pi(k^2,p^2)
=\int_0^{\infty} ds_1\int_0^{\infty} ds_2\;e^{-s_1\sigma_1}e^{-s_2\sigma_2}\;
\rho(s_1,s_2),
\end{eqnarray}
from which we can subtract continuum by cutting the integral at continuum thresholds $s_{01}$ and $s_{02}$. The spectral density $\rho(k^2,p^2)$ can be obtained by taking another double Borel transformations on \eqref{eq:8}:
\begin{eqnarray}
\label{eq:9}
\rho(s_1,s_2)=\mathcal{B}_{-\sigma_1}^{\frac{1}{s_1}}\mathcal{B}_{-\sigma_2}^{\frac{1}{s_2}}\mathcal{B}_{k^2}^{\frac{1}{\sigma_1}}\mathcal{B}_{p^2}^{\frac{1}{\sigma_2}}\Pi(k^2,p^2).
\end{eqnarray}
After invoking the double Borel transformations to the phenomenological representations \eqref{eq:4}, \eqref{eq:5} and \eqref{eq:6}, and compare them with the QCD side \eqref{eq:8} using \eqref{eq:9}, we get the master equations of light-cone QCD sum rules:
\begin{eqnarray}
\label{eq:11}
f^{T}_{b_1}f_{\pi_1}m_{\pi_1}^3g_{b_1}^1 e^{-m_{b_1}^2\sigma_1-m_{\pi_1}^2\sigma_2}
=\int_0^{s_{01}} ds_1\int_0^{s_{02}} ds_2\;e^{-s_1\sigma_1}e^{-s_2\sigma_2}\;
\mathcal{B}_{-\sigma_1}^{\frac{1}{s_1}}\mathcal{B}_{-\sigma_2}^{\frac{1}{s_2}}
\mathcal{B}_{k^2}^{\frac{1}{\sigma_1}}\mathcal{B}_{p^2}^{\frac{1}{\sigma_2}}\Pi_{b1;1}^T(k^2,p^2)\,,
\end{eqnarray}
\begin{eqnarray}
\label{eq:12}
f^{T}_{b_1}f_{\pi_1}m_{\pi_1}^3g_{b_1}^2 e^{-m_{b_1}^2\sigma_1-m_{\pi_1}^2\sigma_2}
=\int_0^{s_{01}} ds_1\int_0^{s_{02}} ds_2\;e^{-s_1\sigma_1}e^{-s_2\sigma_2}\;
\mathcal{B}_{-\sigma_1}^{\frac{1}{s_1}}\mathcal{B}_{-\sigma_2}^{\frac{1}{s_2}}
\mathcal{B}_{k^2}^{\frac{1}{\sigma_1}}\mathcal{B}_{p^2}^{\frac{1}{\sigma_2}}\Pi_{b1;2}^T(k^2,p^2)\,,
\end{eqnarray}
\begin{eqnarray}
\label{eq:13}
f^T_{\rho}f_{\pi_1}m_{\pi_1}^3g_{\rho}e^{-m_{\rho}^2\sigma_1-m_{\pi_1}^2\sigma_2}
=\int_0^{s_{01}} ds_1\int_0^{s_{02}} ds_2\;e^{-s_1\sigma_1}e^{-s_2\sigma_2}\;
\mathcal{B}_{-\sigma_1}^{\frac{1}{s_1}}\mathcal{B}_{-\sigma_2}^{\frac{1}{s_2}}
\mathcal{B}_{k^2}^{\frac{1}{\sigma_1}}\mathcal{B}_{p^2}^{\frac{1}{\sigma_2}}\Pi^T_{\rho}(k^2,p^2)\,,
\end{eqnarray}
\begin{eqnarray}
\label{eq:14}
if_{b_1}f_{\pi_1}m_{\pi_1}^3g_{b_1}^1 e^{-m_{b_1}^2\sigma_1-m_{\pi_1}^2\sigma_2}
=\int_0^{s_{01}} ds_1\int_0^{s_{02}} ds_2\;e^{-s_1\sigma_1}e^{-s_2\sigma_2}\;
\mathcal{B}_{-\sigma_1}^{\frac{1}{s_1}}\mathcal{B}_{-\sigma_2}^{\frac{1}{s_2}}
\mathcal{B}_{k^2}^{\frac{1}{\sigma_1}}\mathcal{B}_{p^2}^{\frac{1}{\sigma_2}}\Pi_{b1;1}^D(k^2,p^2)\,,
\end{eqnarray}
\begin{eqnarray}
\label{eq:15}
if_{b_1}f_{\pi_1}m_{\pi_1}^3g_{b_1}^2 e^{-m_{b_1}^2\sigma_1-m_{\pi_1}^2\sigma_2}
=\int_0^{s_{01}} ds_1\int_0^{s_{02}} ds_2\;e^{-s_1\sigma_1}e^{-s_2\sigma_2}\;
\mathcal{B}_{-\sigma_1}^{\frac{1}{s_1}}\mathcal{B}_{-\sigma_2}^{\frac{1}{s_2}}
\mathcal{B}_{k^2}^{\frac{1}{\sigma_1}}\mathcal{B}_{p^2}^{\frac{1}{\sigma_2}}\Pi_{b1;2}^D(k^2,p^2)\,,
\end{eqnarray}
where contributions from excited states and continuum have been subtracted from both phenomenological and QCD sides.

\section{Light-cone expansion of the correlation functions}
\label{LCexpansion}

We expand the correlation function near the light-cone in distribution amplitudes calculated in \cite{Ball:1998tj}. Contributions of different decay modes mix in the final results. Depending on the certain current used in the correlation function, it is sometimes not quite straightforward to pick out the particular tensor structures corresponding to certain decay modes, which in our case holds for the tensor current. Before presenting our results of light-cone expansion, we show how to separate the tensor structures in the light-cone expansion of correlation functions.

For the tensor current correlation function, the tensors that appear in the final results involve Levi-Civita tensors. Generally we can form six tensor structures with a Levi-Civita tensor and two independent momentums with three independent Lorentz indices $\mu$, $\nu$ and $\alpha$ ($\mu$, $\nu$ are anti-symmetric). They are
\begin{eqnarray}
T_1=\epsilon_{\mu\nu\rho\alpha}p^\rho\,,\,T_2=\epsilon_{\mu\alpha\rho\sigma}k^\rho p^\sigma p_\nu-\epsilon_{\nu\alpha\rho\sigma}k^\rho p^\sigma p_\mu\,,\, T_3=\epsilon_{\mu\nu\rho\alpha}k^\rho\,,\,\\\nonumber T_4=\epsilon_{\mu\nu\rho\sigma}k^\rho p^\sigma k_\alpha\,,\,T_5=\epsilon_{\mu\nu\rho\sigma}k^\rho p^\sigma p_\alpha\,,\,
T_6=\epsilon_{\mu\alpha\rho\sigma}k^\rho p^\sigma k_\nu-\epsilon_{\nu\alpha\rho\sigma}k^\rho p^\sigma k_\mu.
\end{eqnarray}
Actually, only four of the above tensors are independent. One can prove the formula below:
\begin{eqnarray}
\label{eq:17}
T_5-T_2=p^2 T_3-p\cdot k T_1.
\end{eqnarray}
By exchanging $p$ and $k$, we get
\begin{eqnarray}
\label{eq:18}
T_4-T_6=-k^2 T_1+p\cdot k T_3.
\end{eqnarray}
By using \eqref{eq:17} and \eqref{eq:18}, $T_1$ and $T_2$ that appear in the final results of the light-cone expansion can be expressed in terms of $T_3$--$T_6$.
On the phenomenological side of correlation function of the tensor current, tensor structures corresponding to different decay modes are as below:
\begin{eqnarray}
\label{eq:19}
\pi_1\rightarrow b_1\pi\,&:&\,g^1_{b_1}\epsilon_{\mu\nu\rho\sigma}k^\rho(-\frac{ p^\sigma p_\alpha}{p^2}+g_\alpha^{\ \sigma})\qquad\ \ \ \ \ \ \ \ \sim\, -\frac{1}{p^2}T_5+T_3\nonumber\\
&+&g^2_{b_1}\epsilon_{\mu\nu\rho\sigma}k^\rho(-\frac{k\cdot p}{p^2}p^\sigma p_\alpha+p^\sigma k_\alpha)\qquad\ \sim\, -\frac{p\cdot k}{p^2}T_5+T_4\nonumber\\
\pi_1\rightarrow \rho\pi\,&:&\,g_\rho(\epsilon_{\rho\sigma\mu\alpha}k^\rho p^\sigma k_\nu-\epsilon_{\rho\sigma\nu\alpha}k^\rho p^\sigma k_\mu)\qquad\sim\, T_6\nonumber\\
0^{++}\rightarrow b_1\pi\,&:&\,g_{b_1}'\epsilon_{\mu\nu\rho\sigma}k^\rho p^\sigma p_\alpha\qquad\ \ \ \ \ \ \ \ \ \ \ \ \ \ \ \ \ \ \ \ \ \sim\, T_5
\end{eqnarray}

From \eqref{eq:19} we can see that $T_3$, $T_4$ and $T_6$ are the characteristic tensors for $b_1\pi$ and $\rho\pi$ decay modes, of which the corresponding terms on the QCD side can be extracted to compare with the phenomenological side. After doing this, we obtain the QCD side of the tensor current correlation function with light-cone expansion.

The tensor structure for the correlation function of the derivative current are much simpler, we can see from \eqref{eq:6} that $g_{\mu\nu}$ and $p_\mu k_\nu$ can be the characteristic tensors (the $0^{++}$ decay mode has a tensor structure parallel to $p_\nu$ due to $\langle 0|J_\nu^{H}(0)|0^{++}\rangle \sim p_\nu$).

Using the method above, we are able to disentangle the tensor structures and get the following results of light-cone expansion:
\begin{eqnarray}%\label{b1presumrule1}
\label{eq:20}
\mathcal{B}_{k^2}^{\frac{1}{\sigma_1}}\mathcal{B}_{p^2}^{\frac{1}{\sigma_2}}\Pi_{b1;1}^T(k^2,p^2)
=-\frac{\sqrt{2}\pi f_\pi m_\pi^2}{108(m_u+m_d)}\langle \alpha_sG^2\rangle\biggl\lbrace\frac{1}{2}[\phi_\sigma'(u_0)-\phi_\sigma'(\bar{u}_0)]+3[\phi_p(u_0)+\phi_p(\bar{u}_0)]+3(\phi_p^{[u]}+\phi_p^{[\bar{u}]})\biggl\rbrace,
\end{eqnarray}
\begin{eqnarray}%\label{b1presumrule1}
\label{eq:21}
\mathcal{B}_{k^2}^{\frac{1}{\sigma_1}}\mathcal{B}_{p^2}^{\frac{1}{\sigma_2}}\Pi_{b1;2}^T(k^2,p^2)
&=&-\frac{\sqrt{2}f_\pi m_\pi^2}{(m_u+m_d)}(\mathcal{T}^{[\alpha_1]}+\mathcal{T}^{[\alpha_2]})\frac{1}{\sigma}\nonumber\\
&&+\frac{\sqrt{2}\pi f_\pi m_\pi^2}{108(m_u+m_d)}\langle \alpha_sG^2\rangle\biggl\lbrace[\phi_\sigma(u_0)+\phi_\sigma(\bar{u}_0)](\sigma_1-\sigma_2)+6(\phi_p^{[u]}+\phi_p^{[\bar{u}]})\sigma_2\biggl\rbrace,
\end{eqnarray}
\begin{eqnarray}%\label{b1presumrule1}
\label{eq:22}
\mathcal{B}_{k^2}^{\frac{1}{\sigma_1}}\mathcal{B}_{p^2}^{\frac{1}{\sigma_2}}\Pi_{\rho}^T(k^2,p^2)
=\frac{\sqrt{2}\pi f_\pi m_\pi^2}{108(m_u+m_d)}\langle \alpha_sG^2\rangle\biggl\lbrace[\phi_\sigma(u_0)+\phi_\sigma(\bar{u}_0)]\sigma-6(\phi_p^{[u]}+\phi_p^{[\bar{u}]})\sigma_2\biggl\rbrace,
\end{eqnarray}
\begin{eqnarray}%\label{b1presumrule1}
 \label{eq:23}
\mathcal{B}_{k^2}^{\frac{1}{\sigma_1}}\mathcal{B}_{p^2}^{\frac{1}{\sigma_2}}\Pi_{b_1;1}^D(k^2,p^2)
=\frac{i\sqrt{2}\pi f_\pi m_\pi^2}{108(m_u+m_d)}\biggl\lbrace\frac{1}{2}\left[\phi_\sigma'(u_0)-\phi_\sigma'(\bar{u}_0)\right]-3\left[\phi_p(u_0)+\phi_p(\bar{u}_0)\right]\biggl\rbrace \frac{1}{\sigma}\langle \alpha_sG^2\rangle,
\end{eqnarray}
\begin{eqnarray}%\label{b1presumrule2}
\label{eq:24}
&&\mathcal{B}_{k^2}^{\frac{1}{\sigma_1}}\mathcal{B}_{p^2}^{\frac{1}{\sigma_2}}\Pi_{b_1;2}^D(k^2,p^2)\nonumber\\
&&=\frac{i\sqrt{2}\pi f_\pi m_\pi^2}{54(m_u+m_d)}
\biggl\{(1-3u_0-\frac{\bar{u}_0}{u_0})[\phi_\sigma(u_0)+\phi_\sigma(\bar{u}_0)]+u_0\bar{u}_0[\phi_\sigma'(u_0)-\phi_\sigma'(\bar{u}_0)]\nonumber\\
&&\phantom{=}+3u_0(1-u_0)\left[\phi_p(u_0)+\phi_p(\bar{u}_0)\right]\biggl\}\langle\alpha_sG^2\rangle
\nonumber\\
&&\phantom{=}+\frac{i\sqrt{2}f_\pi m_\pi^2}{(m_u+m_d)}\biggl[u_0\mathcal{T}(u_0,\bar{u}_0,0)+u_0\mathcal{T}(\bar{u}_0,u_0,0)
+u_0\left(\frac{\partial\mathcal{T}}{\partial\alpha_3}-\frac{\partial\mathcal{T}}{\partial\alpha_2}\right)^{[\alpha_1]}\nonumber\\
&&\phantom{=}+u_0\left(\frac{\partial\mathcal{T}}{\partial\alpha_3}-\frac{\partial\mathcal{T}}{\partial\alpha_1}\right)^{[\alpha_2]}
-\mathcal{T}^{[\alpha_1]}-\mathcal{T}^{[\alpha_2]}\biggl]\frac{1}{\sigma^2},
\end{eqnarray}
where the Borel variable $\sigma=\sigma_1+\sigma_2$. We have adopted the vacuum saturation approximation and the definitions of the notations can be found in Appendix A. We have used the same definitions of the pion distribution amplitudes of those used in \cite{Huang:2010dc}, which have been calculated in \cite{Ball:1998tj}. We also use the current $J^D_\mu=\bar{d}\overleftrightarrow{D}_\mu\gamma_5u$ instead of $\bar{d}\overleftrightarrow{\partial}_\mu\gamma_5u$ used in \cite{Huang:2010dc}, which lead to discrepancies in the final results of light-cone expansion and contradictory results in the numerical analysis.
We have also compared our results from the non-covariant derivative current with those obtained in \cite{Huang:2010dc},
only finding a misprint: there are extra $u_0$ factors in the $\mathcal{T}^{[\alpha_1]}$ and $\mathcal{T}^{[\alpha_2]}$ terms of the light-cone sum rules for $g_{b_1}^2$ in \cite{Huang:2010dc}.

\section{integrals of the spectral densities}
\label{integral}

After substituting the pion distribution amplitudes with the expressions in Appendix A, $\mathcal{B}_{k^2}^{\frac{1}{\sigma_1}}\mathcal{B}_{p^2}^{\frac{1}{\sigma_2}}\Pi^{T,D}(k^2,p^2)$ in \eqref{eq:11}--\eqref{eq:15} are of three types:  $\frac{\sigma_2^m}{(\sigma_1+\sigma_2)^n}$, $\ln\frac{\sigma_2}{\sigma_1+\sigma_2}$ and $\sigma_2\ln\frac{\sigma_2}{\sigma_1+\sigma_2}$, where $m\geqslant0$ and $n>0$.

For the first type, the general form of the spectral density integral can be calculated in the following procedure:
\begin{eqnarray}
\label{eq:25}
&&\int_0^{s_{01}} ds_1\int_0^{s_{02}} ds_2\;e^{-s_1\sigma_1}e^{-s_2\sigma_2}\;
\mathcal{B}_{-\sigma_1}^{\frac{1}{s_1}}\mathcal{B}_{-\sigma_2}^{\frac{1}{s_2}}
\frac{\sigma_2^m}{(\sigma_1+\sigma_2)^{n}}\nonumber\\
&=&\int_0^{s_{01}} ds_1\int_0^{s_{02}} ds_2\;e^{-s_1\sigma_1}e^{-s_2\sigma_2}\;
\frac{1}{\Gamma(n)}\frac{\partial^m}{\partial s_2^m}\left[\delta(s_1-s_2)s_2^{n-1}\right]\nonumber\\
&=&\int_0^{s_{01}} ds_1\int_0^{s_{02}} ds_2\;e^{-s_1\sigma_1}e^{-s_2\sigma_2}\;
\frac{1}{\Gamma(n)}\frac{\partial^m\delta(s_2-s_1)}{\partial s_2^m}s_1^{n-1},
\end{eqnarray}
where $s_{01}<s_{0_2}$ is a reasonable assumption according to $m_{b_1,\rho}<m_{\pi_1}$. The power of $\frac{\partial}{\partial s_2}$ in the last equation of \eqref{eq:25} can be reduced using integration by parts. Doing this one time, we get the surface term as below:
\begin{eqnarray}
\label{eq:26}
\bigtriangleup=\frac{(-1)^{m}}{\Gamma(n)}\int_0^{s_{01}} ds_1\;e^{-s_1\sigma_1}\;
\frac{\partial^{m-1}\delta(s_1)}{\partial s_1^{m-1}}s_1^{n-1},
\end{eqnarray}
which is well-defined and vanishing only if $n>m$. To avoid the ambiguity arising from the surface term, we shift the lower limit of the integral \eqref{eq:25} by a small constant and get
\begin{eqnarray}
&&\int_{0^+}^{s_{01}} ds_1\int_0^{s_{02}} ds_2\;e^{-s_1\sigma_1}e^{-s_2\sigma_2}\;
\frac{1}{\Gamma(n)}\frac{\partial^m\delta(s_2-s_1)}{\partial s_2^m}s_1^{n-1}\nonumber\\
&=&\frac{\sigma_2^m}{\Gamma(n)}\int_0^{s_{01}} ds_1 e^{-s_1(\sigma_1+\sigma_2)}\;s_1^{n-1}\nonumber\\
&=&\sigma_2^m(\sigma_1+\sigma_2)^{-n}\biggl\{1-\frac{\Gamma[n,(\sigma_1+\sigma_2)s_{01}]}{\Gamma(n)}\biggl\}\nonumber\\
&=&\sigma^m_2(\sigma_1+\sigma_2)^{-n}f_{n-1}[(\sigma_1+\sigma_2)s_{01}]\,,
\end{eqnarray}
where $f_n(x)=1-e^{-x}\sum^n_{i=0}\frac{x^i}{i!}$, $\Gamma[x]$ is the Eular Gamma fuction, and $\Gamma[x,y]$ is the incomplete Gamma function\,.

For the second type of the spectral density integrals, we have
\begin{eqnarray}
&&\int_0^{s_{01}} ds_1\int_0^{s_{02}} ds_2\;e^{-s_1\sigma_1}e^{-s_2\sigma_2}\;
\mathcal{B}_{-\sigma_1}^{\frac{1}{s_1}}\mathcal{B}_{-\sigma_2}^{\frac{1}{s_2}}
\ln\frac{\sigma_2}{\sigma_1+\sigma_2}\nonumber\\
&=&\ln\frac{\sigma_2}{\sigma_1+\sigma_2}-\int_{s_{01}}^{\infty} ds_1\int_{s_{02}}^{\infty} ds_2\;e^{-s_1\sigma_1}e^{-s_2\sigma_2}\;
\mathcal{B}_{-\sigma_1}^{\frac{1}{s_1}}\mathcal{B}_{-\sigma_2}^{\frac{1}{s_2}}\ln\frac{\sigma_2}{\sigma_1+\sigma_2}\nonumber\\
&=&\ln\frac{\sigma_2}{\sigma_1+\sigma_2}-\int_{s_{01}}^{\infty} ds_1\int_{s_{02}}^{\infty} ds_2\;e^{-s_1\sigma_1}e^{-s_2\sigma_2}\;\frac{1}{s_2}\delta(s_1-s_2)\nonumber\\
&=&\ln\frac{\sigma_2}{\sigma_1+\sigma_2}-\Gamma[0,s_{02}(\sigma_1+\sigma_2)].
\end{eqnarray}

Similarly, we get the third type of the spectral density integrals:
\begin{eqnarray}
&&\int_0^{s_{01}} ds_1\int_0^{s_{02}} ds_2\;e^{-s_1\sigma_1}e^{-s_2\sigma_2}\;
\mathcal{B}_{-\sigma_1}^{\frac{1}{s_1}}\mathcal{B}_{-\sigma_2}^{\frac{1}{s_2}}\sigma_2
\ln\frac{\sigma_2}{\sigma_1+\sigma_2}\nonumber\\
&=&\sigma_2\ln\frac{\sigma_2}{\sigma_1+\sigma_2}-\int_{s_{01}}^{\infty} ds_1\int_{s_{02}}^{\infty} ds_2\;e^{-s_1\sigma_1}e^{-s_2\sigma_2}\;
\mathcal{B}_{-\sigma_1}^{\frac{1}{s_1}}\mathcal{B}_{-\sigma_2}^{\frac{1}{s_2}}\ln\frac{\sigma_2}{\sigma_1+\sigma_2}\nonumber\\
&=&\sigma_2\ln\frac{\sigma_2}{\sigma_1+\sigma_2}-\int_{s_{01}}^{\infty} ds_1\int_{s_{02}}^{\infty} ds_2\;e^{-s_1\sigma_1}e^{-s_2\sigma_2}\;\frac{1}{s_1}\frac{d}{ds_2}\delta(s_1-s_2)\nonumber\\
&=&\sigma_2\ln\frac{\sigma_2}{\sigma_1+\sigma_2}+e^{-s_{02}(\sigma_1+\sigma_2)}\frac{1}{s_{02}}-\sigma_2\Gamma[0,s_{02}(\sigma_1+\sigma_2)].
\end{eqnarray}

Using these integral formula, we can transform the master equations \eqref{eq:11}--\eqref{eq:15} into forms of
\begin{gather}
\label{eq:ms1}
g_{b_1}^1=\Pi^T_{b_1;1}(\sigma,s_{01},s_{02})/(f^T_{b_1} f_{\pi_1}m_{\pi_1}^3 e^{-2 m_{b_1}^2 m_{\pi_1}^2/(m_{b_1}^2+m_{\pi_1}^2)\cdot\sigma}),\\
\label{eq:ms2}
g_{b_1}^2=\Pi^T_{b_1;2}(\sigma,s_{01},s_{02})/(f^{T}_{b_1}f_{\pi_1}m_{\pi_1}^3 e^{-2 m_{b_1}^2 m_{\pi_1}^2/(m_{b_1}^2+m_{\pi_1}^2)\cdot\sigma}),\\
\label{eq:ms3}
g_\rho=\Pi^T_\rho(\sigma,s_{01},s_{02})/(f^T_{\rho}f_{\pi_1}m_{\pi_1}^3e^{-2 m_{\rho}^2 m_{\pi_1}^2/(m_{\rho}^2+m_{\pi_1}^2)\cdot\sigma}),\\
\label{eq:ms4}
g_{b_1}^1=\Pi^D_{b_1;1}(\sigma,s_{01})/(if_{b_1}f_{\pi_1}m_{\pi_1}^3 e^{-2 m_{b_1}^2 m_{\pi_1}^2/(m_{b_1}^2+m_{\pi_1}^2)\cdot\sigma}),\\
\label{eq:ms5}
g_{b_1}^2=\Pi^D_{b_1;2}(\sigma,s_{01})/(if_{b_1}f_{\pi_1}m_{\pi_1}^3e^{-2 m_{b_1}^2 m_{\pi_1}^2/(m_{b_1}^2+m_{\pi_1}^2)\cdot\sigma})
\end{gather}
respectively, where $\sigma=\sigma_1$+$\sigma_2$, and we assume $\frac{\sigma_2}{\sigma_1}=\frac{m_{b_1,\rho}^2}{m_{\pi_1}^2}$.

%%%%%%%%%%%%%%%%%%%%%%%%%%%%%%%%%%%%%%%%%%%%%%%%%%%%%%%%%%%%%%%%%%%%%%%%%%%%%%%%%%%%%%%%%%%%%%%%%%%%

\section{Results and Discussions}
\label{numerical}

To obtain predictions for $g_{b_1}^1$, $g_{b_1}^2$ and $g_\rho$ from the master
equations \eqref{eq:ms1}--\eqref{eq:ms5},
we vary the continuum thresholds $s_{01}$ and $s_{02}$ within the physically acceptable ranges to find the stable regions for the couplings, in which the dependence of the couplings on $\sigma$ is weak , which allows theoretical predictions.
Since there is still different possibilities for the mass of $1^{-+}$
hybrid, we consider three different values of the hybrid mass, i.e., $m_{\pi_1}$ = 1.6\,GeV, 1.8\,GeV and 2.0\,GeV, and we use the decay constant $f_{\pi_1}=0.025\,\textrm{GeV}$ deduced from QCD sum rules \cite{Narison:2009vj,Huang:2014hya}.

\subsection{Numerical analysis for $g_{b_1}^1$}

\begin{figure}[htbp]
\centering
\includegraphics[scale=0.62]{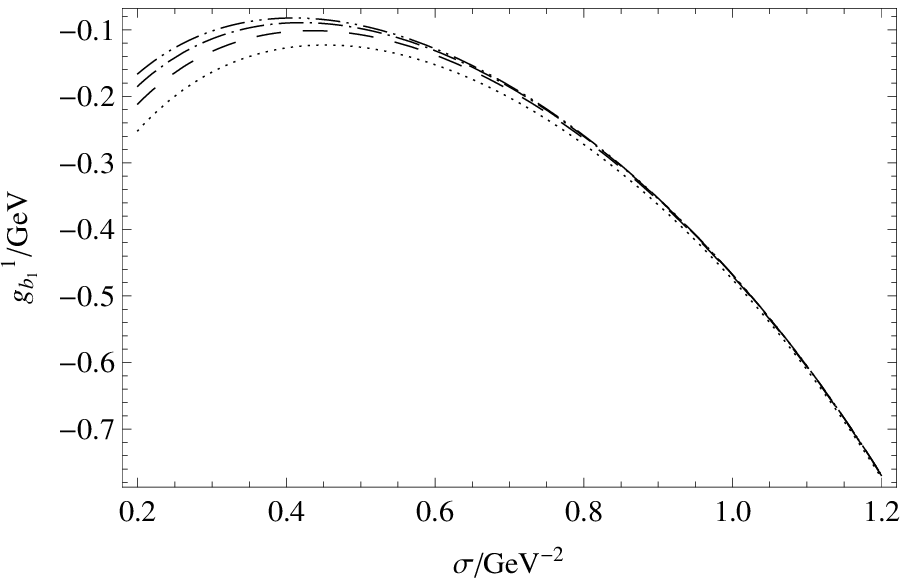}
\includegraphics[scale=0.62]{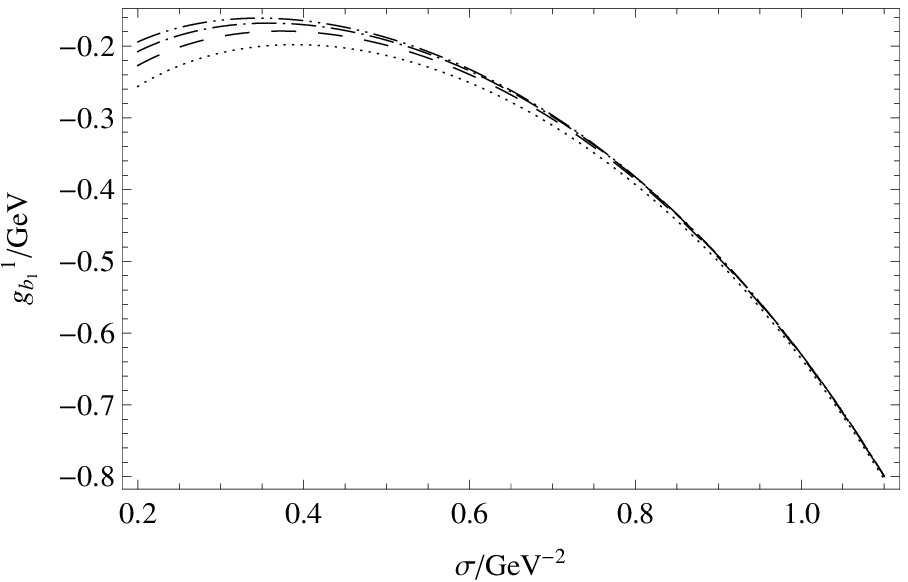}
\includegraphics[scale=0.62]{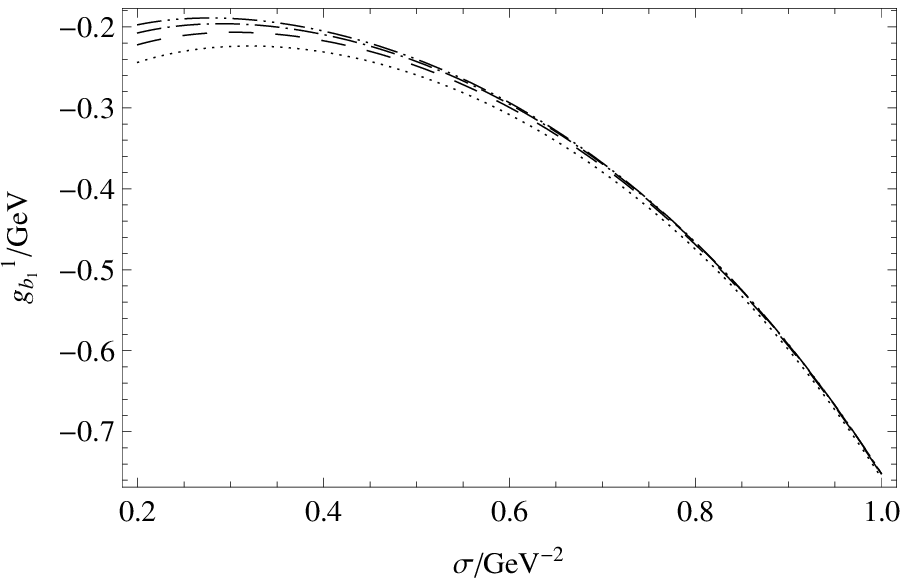}
\caption{\label{fig:g11T} $g_{b_1}^1-\sigma$ curves from the master
equation \eqref{eq:ms1} for $m_{\pi_1}$ = 1.6\,GeV, $m_{\pi_1}$ = 1.8\,GeV, $m_{\pi_1}$ = 2.0\,GeV. The dotted
line, the dashed line, dot-dashed line and the dot-dot-dashed line
denote $\{s_{01}, s_{02}\}$ = \{3\,GeV$^2$, 4\,GeV$^2$\}, \{3\,GeV$^2$,
5\,GeV$^2$\}, \{3\,GeV$^2$, 6\,GeV$^2$\} and \{3\,GeV$^2$, 7\,GeV$^2$\} respectively.
}
\end{figure}

\begin{figure}[htbp]
\centering
\includegraphics[scale=0.7]{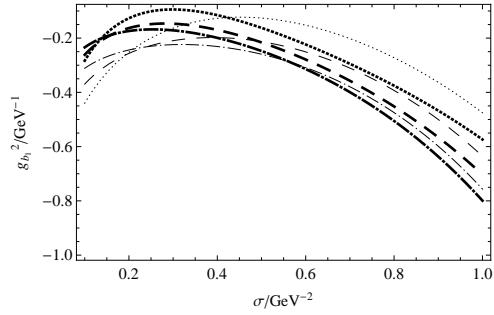}
\caption{\label{fig:g12T} $g_{b_1}^1-\sigma$ curves from the master
equation \eqref{eq:ms1}. The dotted lines, the dashed lines and
the dot-dashed lines denote $m_{\pi_1}$ = 1.6\,GeV, 1.8\,GeV and
2.0\,GeV respectively. All thick lines denote $\{s_{01}, s_{02}\}$ = \{5\,GeV$^2$, 6\,GeV$^2$\}
while the other lines denote $\{s_{01}, s_{02}\}$ = \{3\,GeV$^2$, 4\,GeV$^2$\}.
}
\end{figure}

We first consider the master equation \eqref{eq:ms1}. Numerically we use $m_{b_1}=1.235\,\textrm{GeV}$ and $f^T_{b_1}(2~\textrm{GeV})=0.18\,\textrm{GeV}$ in \cite{Jansen:2009yh}.
There are two continuum thresholds $s_{01}$ and $s_{02}$
in this sum rule, which seems tricky to deal with. However, we find under $s_{02}>s_{01}$ (given $m_{\pi_1}>m_{b_1,\rho}$), $g_{b_1}^1$ depends weakly on  $s_{02}$  (see Figure \ref{fig:g11T}),
which enter the sum rules with the incomplete Gamma function.
Thus for simplicity, we will set $s_{02}=s_{01}+1.0\,\textrm{GeV}^2$ in this sum rule.
By varying the value of $s_{01}$, we can observe how the
$g_{b_1}^1-\sigma$ curves change. In principle, we expect $g_{b_1}$ depend weekly on the external parameters ($\sigma$,$s_{01}$), which has been emphasized in
traditional QCD sum rules \cite{Narison:2002pw}. In practice, we find $g_{b_1}$ shows stability in $\sigma$ (by the extreme values in Figure \ref{fig:g12T}), but no
stability in $s_{01}$. In fact, $g_{b_1}$ increases gradually with $s_{01}$, which means $s_{01}$ cannot be fixed from the stability criterion. Therefore it is
appropriate to consider a conservative range of $g_{b_1}$ by varying $s_{01}$ within its physically acceptable range (where $\sigma$ stability should also be ensured). In Figure \ref{fig:g12T}, we plot the optimal results obtained in the region $s_{01}=3\sim5\,\textrm{GeV}^2$.
By reading the extremum values for $g_{b_1}^1$ from the curves, we can obtain estimated
values for $g_{b_1}^1$. For $m_{\pi_1}=1.6\,\textrm{GeV}$, we
find $g_{b_1}^1=-0.12\sim -0.09\,\textrm{GeV}$. If the mass of $1^{-+}$ hybrid
is larger than 1.6\,GeV, we will obtain different values of $g_{b_1}^1$. We
find $g_{b_1}^1=-0.20\sim-0.15\,\textrm{GeV}$ for $m_{\pi_1}=1.8\,\textrm{GeV}$
and $g_{b_1}^1=-0.22\sim-0.17\,\textrm{GeV}$ for $m_{\pi_1}=2.0\,\textrm{GeV}$.


\begin{figure}[htbp]
\centering
\includegraphics[scale=0.7]{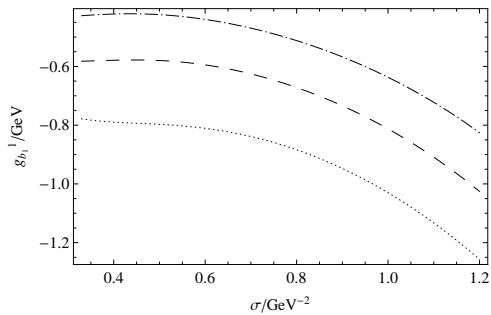}
\caption{\label{fig:g1D} $g_{b_1}^1-\sigma$ curve from master equation
\eqref{eq:ms4}. The dotted line, the dashed line and the dot-dashed
line denote \{$s_{01},m_{\pi_1}$\}=\{7\,\textrm{GeV}$^2$, 1.6\,\textrm{GeV}\},
\{9\,\textrm{GeV}$^2$, 1.8\,\textrm{GeV}\} and \{11\,\textrm{GeV}$^2$, 2.0\,\textrm{GeV}\} respectively.}
\end{figure}

The sum rule for the derivative current \eqref{eq:ms4}
provides a second way to estimate the value of $g_{b_1}^1$, for which we use $f_{b_1}(2~\textrm{GeV})=0.18\,\textrm{GeV}$ from \cite{Reinders:1984sr}.
In this sum rule, there is only one continuum threshold
$s_{01}$. However, this sum rule does not reach stability in $\sigma$
unless we use large values of $s_{01}$.
In Figure \ref{fig:g1D}, we plot curves where stability in $\sigma$ is initially reached as we increase $s_{01}$
, from which we can read the extreme values of $g_{b_1}^1$, i.e., $g_{b_1}^1=-0.8\,\textrm{GeV}$, $-0.58$\,GeV
and $-0.42$\,GeV for $m_{\pi_1}$ = 1.6\,GeV, 1.8\,GeV and 2.0\,GeV respectively.
Given that the related $s_{01}$ here lies too far away from the square of the ground state mass, we consider values of $g^1_{b_1}$ from the tensor
current LCSR as more reliable predictions.

\subsection{Numerical analysis for $g_{b_1}^2$}

\begin{figure}[htbp]
\centering
\includegraphics[scale=0.7]{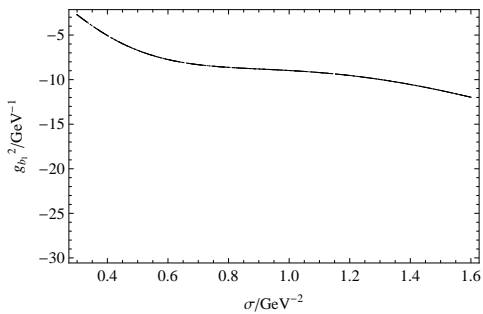}
\caption{\label{fig:g21T} $g_{b_1}^2-\sigma$ curves from master
equation \eqref{eq:ms2} for $m_{\pi_1}$ = 1.6\,GeV. The dotted
line, the dashed line, dot-dashed line and the dot-dot-dashed line
denote $\{s_{01}, s_{02}\}$ = \{5\,GeV$^2$, 6\,GeV$^2$\}, \{5\,GeV$^2$,
7\,GeV$^2$\}, \{5\,GeV$^2$, 8\,GeV$^2$\} and \{5\,GeV$^2$, 9\,GeV$^2$\} respectively.
}
\end{figure}

To obtain the prediction for $g_{b_1}^2$, we first consider the sum rules
for the tensor current. By varying $s_{01}$ and $s_{02}$, we find
$g_{b_1}^2$ is almost insensitive to the value of $s_{02}$. As can be seen in Figure \ref{fig:g21T}, curves corresponding to the same $s_{01}$ and different $s_{02}$ almost overlap with each other. Thus we can still set $s_{02}=s_{01}+1.0\,\textrm{GeV}^2$ in this sum rule.

\begin{figure}[htbp]
\centering
\includegraphics[scale=0.62]{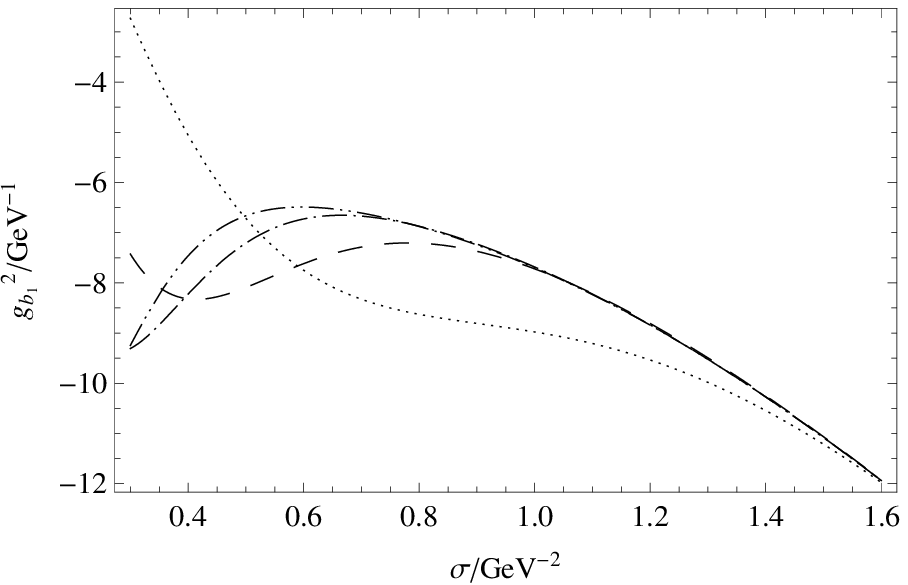}
\includegraphics[scale=0.62]{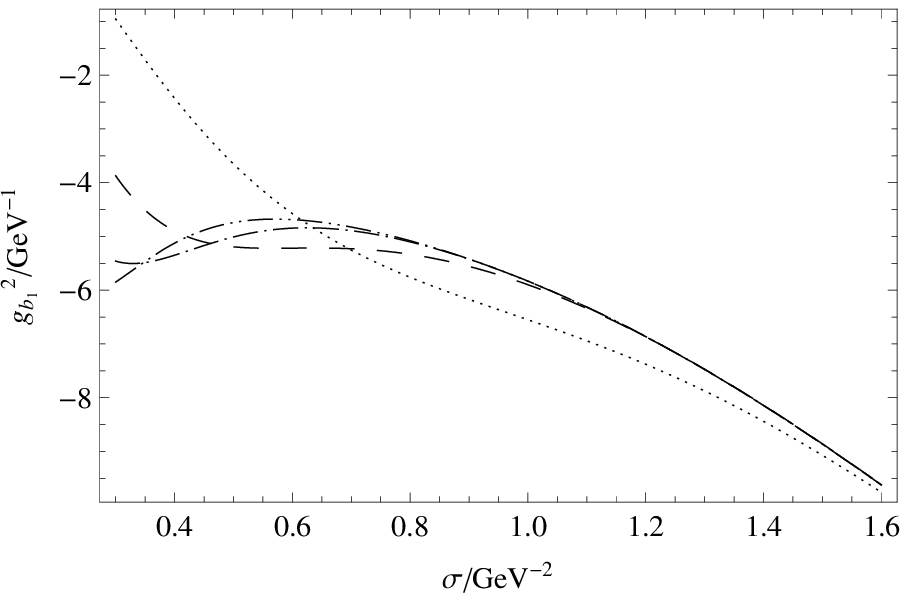}
\includegraphics[scale=0.62]{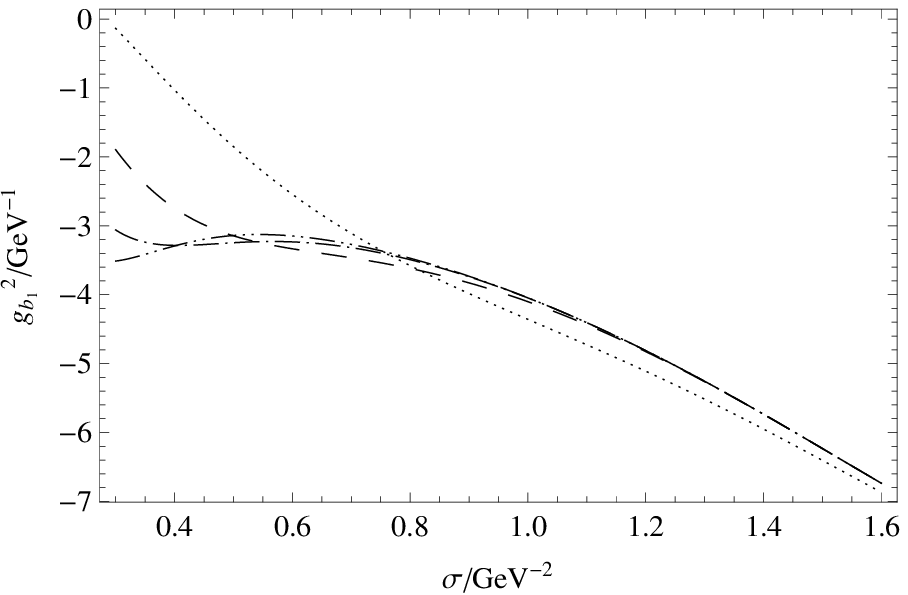}
\caption{\label{fig:g22T} $g_{b_1}^2-\sigma$ curves from master equation
\eqref{eq:ms2} with $m_{\pi_1}$=1.6, 1.8 and 2.0\,$\textrm{GeV}$. The dotted line,
the dashed line, the dot-dashed line and the dot-dot-dashed line denote
$\{s_{01}, s_{02}\}$ = \{5\,GeV$^2$, 6\,GeV$^2$\}, \{8\,GeV$^2$, 9\,GeV$^2$\},
 \{11\,GeV$^2$, 12\,GeV$^2$\} and \{14\,GeV$^2$, 15\,GeV$^2$\} respectively.
 }
\end{figure}

In Figure \ref{fig:g22T}, we can observe how the shape of curves
change when we increase the value of $s_{01}$. The curves are monotonous at low $s_{01}$. If we increase $s_{01}$, the curves will reach stability in $\sigma$.
But even as the stability is initially reached,
the corresponding $s_{01}$ (=7, 8, 10\,GeV$^2$ respectively for  $m_{\pi_1}$ = 1.6, 1.8 and 2.0\,GeV) seems too large for the $b_1$ meson.
Therefore we do not intend to extract specific predictions for $g_{b1}^2$ from LCSR with the
tensor current.

\begin{figure}[!htbp]
\centering
\includegraphics[scale=0.7]{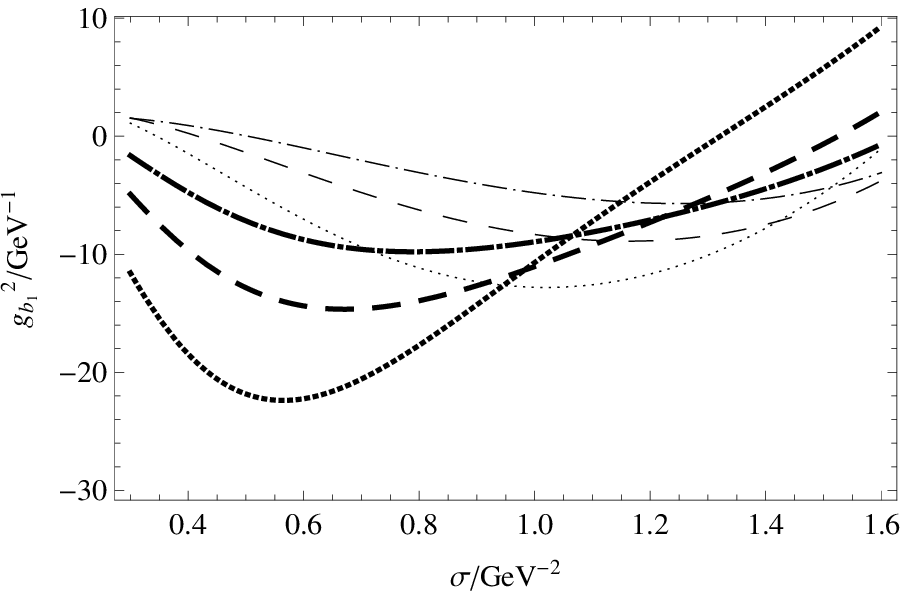}
\caption{\label{fig:g2D} $g_{b_1}^2-\sigma$ curves from master
equation \eqref{eq:ms5}. The dotted lines, the dashed lines and
the dot-dashed lines denote $m_{\pi_1}$ = 1.6\,GeV, 1.8\,GeV and 2.0\,GeV
respectively. All thick lines denote $s_{01}= 5\,\textrm{GeV}^2$ while
the other lines denote $s_{01}= 3\,\textrm{GeV}^2$.
}
\end{figure}
However, the extreme values of the curves will not increase if the
value of $s_{01}$ has reached a ``huge value'', e.g., 14\,GeV$^2$ for $m_{\pi_1}$ = 1.6\,GeV.
which means we can obtain an upper bound of $g_{b_1}^2$.
 In Figure
\ref{fig:g22T}, we obtain $g_{b_1}^2<-6.5\,\textrm{GeV}^{-1}$, $<-4.5$\,GeV$^{-1}$
and $<-3$\,GeV$^{-1}$ for $m_{\pi_1}$ = 1.6\,GeV, 1.8\,GeV and 2.0\,GeV respectively.

The upper bounds above can be compared with the  predictions of $g_{b_1}^2$ from using the derivative current, which are
obtained from the stability criterion in the region $s_{01}=3\sim5\,\textrm{GeV}^2$. We plot all curves in
Figure \ref{fig:g2D}, from which we read $g_{b_1}^2=-12.8\sim -22.4\,\textrm{GeV}^{-1}$,
$-8.9\sim-14.7$\,GeV$^{-1}$ and $-5.7\sim-9.8$\,GeV$^{-1}$ for $m_{\pi_1}$ = 1.6\,GeV,
1.8\,GeV and 2.0\,GeV respectively.

\subsection{Numerical analysis for $g_\rho$}

\begin{figure}[htbp]
\centering
\includegraphics[scale=0.62]{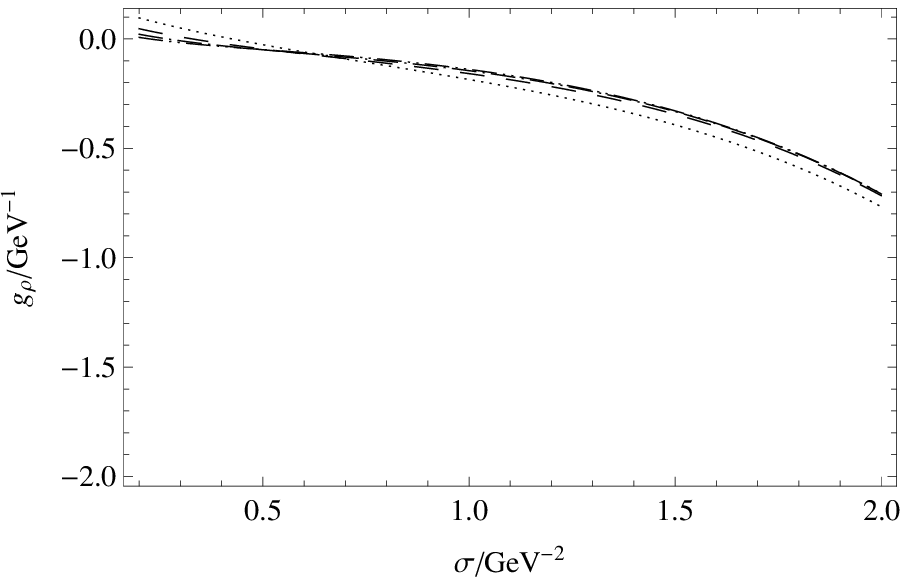}
\includegraphics[scale=0.62]{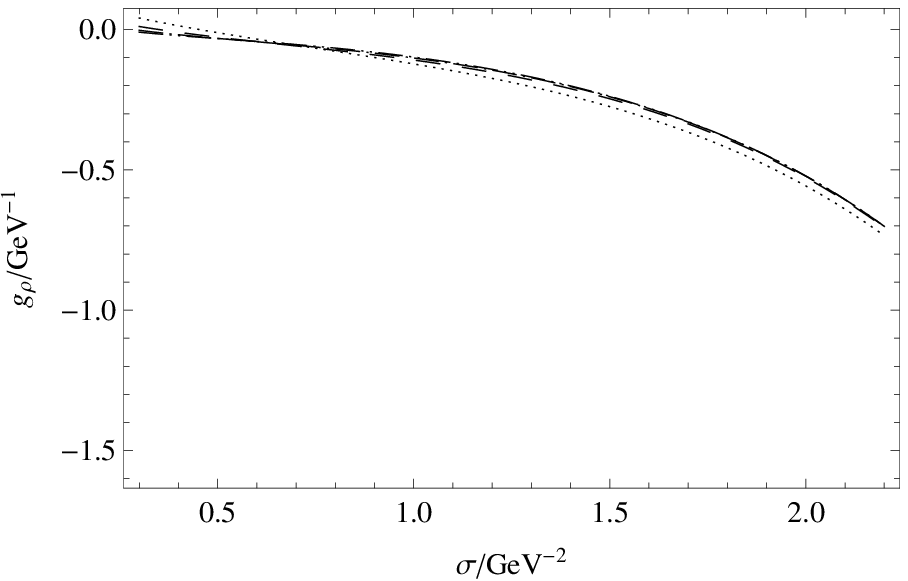}
\includegraphics[scale=0.62]{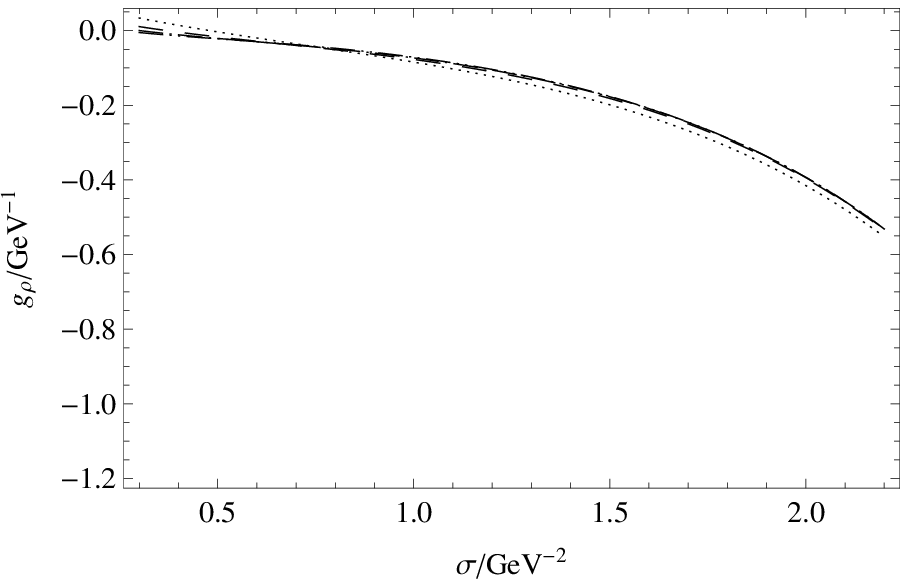}
\caption{\label{fig:grho} $g_\rho-\sigma$ curve for $m_{\pi_1}$ = 1.6\,GeV,
1.8\,GeV and 2.0\,GeV, from master equation \eqref{eq:ms3}.
The dotted line, the dashed line, the dot-dashed line and the dot-dot-dashed line
denote $\{s_{01}, s_{02}\}$ = \{2\,GeV$^{2}$, 3\,GeV$^2$\}, \{3\,GeV$^{2}$, 4\,GeV$^2$\},
 \{4\,GeV$^{2}$, 5\,GeV$^2$\} and \{5\,GeV$^{2}$, 6\,GeV$^2$\} respectively.
 }
\end{figure}

By using sum rule for tensor current, we can also try to obtain the
prediction for $g_\rho$. Numerically we adopt $m_\rho=0.77\,\textrm{GeV}$ and $f^T_\rho(2~\textrm{GeV})=0.159\,\textrm{GeV}$ \cite{Jansen:2009hr,Bakulev:1999gf}. Again the coupling is insensitive to the variation of $s_{02}$ when $s_{01}$ is fixed, and we still assume $s_{02}=s_{01}+1.0\,\textrm{GeV}^2$. As shown in Figure \ref{fig:grho}, although the sum rules for \eqref{eq:ms3} do not reach exact stability in $\sigma$,  in the region where the curves are close to stabilizing, there are intersection
 points for curves with different ($s_{01}$,$s_{02}$). Near these intersection
 points, $g_\rho$ depends weakly on the variation of ($s_{01}$,$s_{02}$), which fulfills the $s_0$ stability criterion of which the importance has been emphasized in traditional QCD sum rules \cite{Narison:2002pw}. Taking the value of $g_\rho$
 at the intersection points, we obtain $g_\rho=-0.06, -0.05\ and -0.06 \,\textrm{GeV}^{-1}$ respectively for 1.6, 1.8 and 2.0\,$\textrm{GeV}$, which suggest $g_\rho$ to be small.

\subsection{Decay widths for $\pi_1\to b_1\pi$ and $\pi_1 \to \rho \pi$}

In the previous subsections, we have obtained values and
ranges of $g_{b_1}^1$ and $g_{b_1}^2$, from both the tensor current
LCSR and derivative current LCSR, and we have also obtained
estimates of $g_\rho$ from tensor current LCSR.

Using these values of $g_{b_1}^1$, $g_{b_1}^2$ and $g_\rho$ as our input
parameters, we can calculate the decay widths for $\pi_1\to b_1\pi$ and $\pi_1 \to \rho\pi$ by using
\begin{equation}
\Gamma(\pi_1\to b_1^+\pi^-+ b_1^-\pi^+)=\frac{1}{12\pi m_{\pi_1}^2}\cdot \left[(g_{b_1}^1)^2
\left(3+\frac{k_{b_1}^2}{m_{b_1}^2}\right)k_{b_1}+2 g_{b_1}^1 g_{b_1}^2
\frac{m_{\pi_1}}{m_{b_1}^2} \sqrt{m_{b_1}^2+k_{b_1}^2} k_{b_1}^3+
(g_{b_1}^2)^2 \frac{m_{\pi_1}^2}{m_{b_1}^2} k_{b_1}^5\right],
\end{equation}
and
\begin{equation}
\Gamma(\pi_1\to \rho^+\pi^-+\rho^-\pi^+)=\frac{g_\rho^2}{6\pi}k_{\rho}^3,
\end{equation}
respectively, where $k_{b_1/\rho}=\sqrt{[(m_{b_1/\rho}-m_{\pi_1})^2-m_\pi^2]
\cdot [(m_{b_1/\rho}+m_{\pi_1})^2-m_\pi^2]}/(2 m_{\pi_1})$.
From these expressions,
we obtain possible values and
lower bounds of $\Gamma(\pi_1\to b_1\pi)$, which are
listed in Table \ref{tab:width}, from which we can see the predictions from $j^D$ LCSR
differs from the very small results obtained in \cite{Huang:2010dc}. This discrepancy is mainly due to our addition of the DAs (distribution amplitudes)
contribution from the covariant derivative of the current $\bar{\psi}\overleftrightarrow{D}_\mu\gamma_5\psi$. Since the light-cone expansion is only known to lower twist, inclusion of any contribution is possible to influence the sum rules to a large extent.
From the last few subsections, we know that some of the sum rules
may suffer from lack of higher twist DAs and are not stable enough within physically acceptable ranges of continuum thresholds, which cause
uncertainties in the predictions. From previous analyses, the best sum rules are $g_{b_1}^1$ from $j^T$ LCSR and $g_{b_1}^2$ from $j^D$ LCSR. Therefore we consider the predictions from these sum rules as the most reliable in our calculation. The decay widths are then $\Gamma(\pi_1\to b_1\pi)$=8--23, 32--86\,MeV and 52--151 for $m_{\pi_1}$=1.6, 1.8 and 2.0\,GeV, which to some extent support the findings from the flux tube model \cite{Kokoski:1985is,Isgur:1985vy,Page:1998gz,Swanson:1997wy} and the modified lattice result \cite{Burns:2006wz}.

\begin{table}[htbp]
\centering
\begin{tabular}{|c|c|c|c|}
\hline
%\hline
\multirow{2}{*}{}& $m_{\pi_1}$=1.6\,GeV& $m_{\pi_1}$=1.8\,GeV& $m_{\pi_1}$=2.0\,GeV\\
\cline{2-4}
&\multicolumn{3}{|c|}{$\Gamma(\pi_1\to b_1\pi)$/MeV}
\\
\hline
$g_{b_1}^1,g_{b_1}^2$ from $j^T$ LCSR&$>2$&$>9$&$>16$\\
\hline
$g_{b_1}^1,g_{b_1}^2$ from $j^D$ LCSR& 20--40 & 46--103 & 62--163 \\
\hline
$g_{b_1}^1$ from $j^T$ LCSR, $g_{b_1}^2$ from $j^D$ LCSR & 8--23 & 32--86 & 52--151 \\
\hline
$g_{b_1}^1$ from $j^D$ LCSR, $g_{b_1}^2$ from $j^T$ LCSR&$>12$ & $>18$ &$>22$\\
\hline
\end{tabular}
\caption{\label{tab:width} Decay widths for $\pi_1 \to b_1\pi$.}
\end{table}

Using the $g_{\rho}$ obtained in the last subsection, we obtain $\Gamma(\pi_1\to \rho\pi)$ = 0.021, 0.037 and 0.040\,MeV for $m_{\pi_1}$ = 1.6, 1.8, 2.0\,GeV, suggesting a small $\rho\pi$ decay width.

\section{Summary and conclusions}
\label{summary}

We have studied the partial decay widths for decay modes $\pi_1\to b_1\pi$ and $\pi_1\to \rho\pi$ using the light-cone QCD sum rules. We use both
the tensor current $\bar\psi\sigma_{\mu\nu}\psi$ and the derivative current $\bar{\psi}\overleftrightarrow{D}_\mu\gamma_5\psi$
as interpolating currents in our calculation.

For the $b_1\pi$ decay mode, we find consistent numerical results (within the errors) of the coupling constants from the sum rules with different interpolating currents. We obtain the partial decay width $\Gamma(\pi_1\to b_1\pi)$= 8--23, 32--86 and 52--151\,MeV for $m_{1^{-+}}$ = 1.6, 1.8 and 2.0\,GeV respectively from the most reliable sum rules, which
provide
support for the flux tube model predictions \cite{Kokoski:1985is,Isgur:1985vy,Page:1998gz,Swanson:1997wy} and the modified lattice predictions \cite{Burns:2006wz}. These results support the hybrid explanations for $\pi_1(1600)$ and $\pi_1(2015)$, both of which have been observed in the $b_1\pi$ channels.

For the $\rho\pi$ decay mode, we have obtained tiny values of the decay widths, which is quite different from the sum rules obtained by using the vector current $\bar \psi\gamma_\mu \psi$. A similar situation also occurs in the sum rules for $\rho$ mass \cite{reinder}. The authors of \cite{reinder} attribute this difference to  two possible reasons: violation of factorization in estimate of four-quark condensate or weak coupling of the tensor current to the $\rho$ meson. Since the value of the $\rho$ meson decay constant  for tensor current obtained from lattice calculation \cite{Jansen:2009hr} is in a reasonable region, we are inclined towards the first reason. Our results go in line with the predictions obtained from the flux tube model \cite{Isgur:1985vy,Page:1998gz,Swanson:1997wy,Close:1994hc}. Since the existence of the $\rho\pi$ decay mode is also uncertain for both $\pi_1(1600)$ and $\pi_1(2015)$ in the experiments \cite{Agashe:2014kda,Meyer:2015eta},
follow-up studies of this decay mode will be of great help for understanding the nature of these exotic states.

As shown from our calculation, higher twist (in our case, twist-5) DAs contributions  may play an important role in stabilizing the sum rules. However, these high twist distribution amplitudes have not been calculated yet. More solid conclusions await the inclusion of contributions from higher twist DAs in the correlation functions.

\begin{acknowledgments}
This work is supported by NSFC under grant 11175153, 11205093 and 11347020, and supported by K. C. Wong Magna Fund in Ningbo University. TGS is supported by the Natural Sciences and Engineering Research Council of Canada (NSERC). Z.R. Huang thanks the University of Saskatchewan for its hospitality.
\end{acknowledgments}

%\clearpage

\begin{appendix}

\section*{Appendix A: Definitions of Pion Distribution Amplitudes and other notations}\label{appendix}

The twist-3 light-cone distribution amplitudes of pion $\phi_{p}(u)$, $\phi_{\sigma}(u)$ and ${\cal T}(\alpha_d,\alpha_u,\alpha_g)$ calculated in \cite{Ball:1998tj} are listed below:
\begin{eqnarray}
\langle 0 | \bar u(z) i\gamma_5 d(-z) | \pi(P)\rangle & = &
\frac{f_\pi m_\pi^2}{m_u+m_d}\, \int_0^1 du \, e^{i(2u-1) pz}\,
\phi_{p}(u)\,,
\label{eq:29}\\
\langle 0 | \bar u(z) \sigma_{\alpha\beta}\gamma_5 d(-z) |
\pi(P)\rangle & = &-\frac{i}{3}\, \frac{f_\pi
  m_\pi^2}{m_u+m_d}  (p_\alpha z_\beta-
p_\beta z_\alpha) \int_0^1 du \, e^{i(2u-1) pz}\,\phi_{\sigma}(u)\,,
\label{eq:30}\\
\langle 0 | \bar u(z) \sigma_{\mu\nu}\gamma_5
  g_sG_{\alpha\beta}(vz) d(-z)| \pi^-(P)\rangle
& = & i\,\frac{f_\pi m_\pi^2}{m_u+m_d} \left(p_\alpha p_\mu
  g_{\nu\beta}^\perp - p_\alpha p_\nu
  g_{\mu\beta}^\perp - p_\beta p_\mu g_{\nu\alpha}^\perp + p_\beta
  p_\nu g_{\alpha\mu}^\perp \right)\nonumber\\ &&\int {\cal D}\underline{\alpha} \, e^{-ipz(\alpha_u
  -\alpha_d + v\alpha_g)} {\cal T}(\alpha_d,\alpha_u,\alpha_g)\,,\label{eq:31}
\end{eqnarray}
where the first two DAs are normalized to unity: $\int_0^1 du\, \phi_{(p,\sigma)}(u) = 1$, and
the projector onto the directions orthogonal to $p$ and $x$ is defined as:
\begin{equation}
g_{\mu\nu}^\perp=g_{\mu\nu}-\frac{1}{pz}(p_{\mu}z_{\nu}+p_{\nu}z_{\mu})\,,
\end{equation}
the integration measure is defined as:
\begin{equation}
\int {\cal D}\underline{\alpha} = \int_0^1 d\alpha_d d\alpha_u
d\alpha_g \delta(1-\alpha_u-\alpha_d-\alpha_g)\,.
\end{equation}

The explicit expressions for the DAs calculated in \cite{Ball:1998tj} are:
\begin{eqnarray}
\phi_p(u)     &=& 1 + \left(30\eta_3 -\frac{5}{2}\, \rho_\pi^2\right)
C_2^{1/2}(\xi) + \left(- 3 \eta_3 \omega_3-\frac{27}{20}\, \rho_\pi^2
  - \frac{81}{10}\, \rho_\pi^2 a_2\right)  C_4^{1/2}(\xi)\,,\\
\phi_\sigma(u)&=& 6u(1-u) \left\{1 + \left(5\eta_3 -\frac{1}{2}\,\eta_3\omega_3 - \frac{7}{20}\,
                  \rho_\pi^2 - \frac{3}{5}\,\rho_\pi^2 a_2 \right) C_2^{3/2}(\xi)\right\}\,,\\
{\cal T}(\underline{\alpha}) &=& 360\eta_3 \alpha_u\alpha_d\alpha_g^2
\left\{1+\omega_3\, \frac{1}{2}\left( 7\alpha_g-3\right)\right\}\,,
\end{eqnarray}
where $\xi=2u-1$ and $C^m_n(\xi)$ are Gegenbauer polynomials. %And the input parameters are as bellow (see \cite{Ball:1998tj}):

%\begin{eqnarray}
%&&m_\pi^2/(m_u+m_d)=(1.6\pm0.2)\,\textrm{GeV},\ \ \ m_{\pi}=0.140\,\textrm{GeV},\ \ \ \rho_\pi^2 \equiv (m_u+m_d)^2/m_\pi^2 \sim O(m_\pi^2),\ \ \ f_\pi=0.131\,\textrm{GeV},\nonumber\\
%&&a_2=0.44,\ \ \eta_3=0.015,\ \ \omega_3=-3, \nonumber
%\end{eqnarray}
%where the renormalization scale is 1\,GeV.
Numerically, We use the following values of the light quark masses and the input parameters involved in the light-cone expansion (at $\mu=1$ GeV) \cite{Ball:1998tj,Narison:2014wqa}:
%\cite{Reinders:1984sr}
\begin{gather*}
m_\pi^2/(m_u+m_d)=(1.6\pm0.2)\,\textrm{GeV},~~m_{\pi}=0.134\,\textrm{GeV},~~\rho_\pi^2 \equiv (m_u+m_d)^2/m_\pi^2 \sim O(m_\pi^2),~~f_\pi=0.131\,\textrm{GeV},\\
a_2=0.44,~~\eta_3=0.015,~~\omega_3=-3,~~\langle\alpha_s G^2\rangle=0.07\,\textrm{GeV}^4.
\end{gather*}\label{decayconstant}
Some other notations that enter in \eqref{eq:20}--\eqref{eq:24} are defined as follows:
\begin{eqnarray}
 u_0=\frac{\sigma_2}{\sigma_1+\sigma_2},\ \ \ \ \ \ \sigma=\sigma_1+\sigma_2,\ \ \ \ \ \bar{u}_0=1-u_0,\ \ \ \ \ \ \bar{u}=1-u,
\end{eqnarray}
\begin{eqnarray}
\phi^{[u]}&=&\int_{u_0}^1\phi(u)\frac{1}{u}\,du\,,\ \ \phi^{[\bar{u}]}=\int_{u_0}^1\phi(\bar{u})\frac{1}{u}\,du\,,\nonumber\\
\mathcal{T}^{[\alpha_1]}&=&\int_0^{\bar{u}_0}\mathcal{T}(\alpha_1,u_0,\bar{u}_0-\alpha_1)\,d\alpha_1\,,\nonumber\\
\mathcal{T}^{[\alpha_2]}&=&\int_0^{\bar{u}_0}\mathcal{T}(u_0,\alpha_2,\bar{u}_0-\alpha_2)\,d\alpha_2\,.
\end{eqnarray}

\end{appendix}


\clearpage

\begin{thebibliography}{20}

%\cite{Agashe:2014kda}
\bibitem{Agashe:2014kda}
  K.~A.~Olive {\it et al.} [Particle Data Group Collaboration],
  %``Review of Particle Physics,''
  Chin.\ Phys.\ C {\bf 38}, 090001 (2014).
%  doi:10.1088/1674-1137/38/9/090001
  %%CITATION = doi:10.1088/1674-1137/38/9/090001;%%
  %3666 citations counted in INSPIRE as of 29 Apr 2016

%\cite{Meyer:2015eta}
\bibitem{Meyer:2015eta}
  C.~A.~Meyer and E.~S.~Swanson,
  %``Hybrid Mesons,''
  Prog.\ Part.\ Nucl.\ Phys.\  {\bf 82}, 21 (2015)
%  doi:10.1016/j.ppnp.2015.03.001
  [arXiv:1502.07276 [hep-ph]].
  %%CITATION = doi:10.1016/j.ppnp.2015.03.001;%%
  %11 citations counted in INSPIRE as of 29 Apr 2016

  %\cite{Chanowitz:1982qj}
\bibitem{Chanowitz:1982qj}
  M.~S.~Chanowitz and S.~R.~Sharpe,
  %``Hybrids: Mixed States of Quarks and Gluons,''
  Nucl.\ Phys.\ B {\bf 222}, 211 (1983)
  Erratum: [Nucl.\ Phys.\ B {\bf 228}, 588 (1983)].
%  doi:10.1016/0550-3213(83)90635-1
  %%CITATION = doi:10.1016/0550-3213(83)90635-1;%%
  %312 citations counted in INSPIRE as of 29 Apr 2016

%\cite{Barnes:1982tx}
\bibitem{Barnes:1982tx}
  T.~Barnes, F.~E.~Close, F.~de Viron and J.~Weyers,
  %``Q anti-Q G Hermaphrodite Mesons in the MIT Bag Model,''
  Nucl.\ Phys.\ B {\bf 224}, 241 (1983).
%  doi:10.1016/0550-3213(83)90004-4
  %%CITATION = doi:10.1016/0550-3213(83)90004-4;%%
  %241 citations counted in INSPIRE as of 29 Apr 2016

%\cite{Lacock:1998be}
\bibitem{Lacock:1998be}
  P.~Lacock {\it et al.} [TXL Collaboration],
  %``Hybrid and orbitally excited mesons in full QCD,''
  Nucl.\ Phys.\ Proc.\ Suppl.\  {\bf 73}, 261 (1999)
%  doi:10.1016/S0920-5632(99)85042-7
  [hep-lat/9809022].
  %%CITATION = doi:10.1016/S0920-5632(99)85042-7;%%
  %80 citations counted in INSPIRE as of 29 Apr 2016

%\cite{McNeile:1998cp}
\bibitem{McNeile:1998cp}
  C.~McNeile {\it et al.},
  %``Exotic meson spectroscopy from the clover action at beta = 5.85 and beta = 6.15,''
  Nucl.\ Phys.\ Proc.\ Suppl.\  {\bf 73}, 264 (1999)
%  doi:10.1016/S0920-5632(99)85043-9
  [hep-lat/9809087].
  %%CITATION = doi:10.1016/S0920-5632(99)85043-9;%%
  %58 citations counted in INSPIRE as of 29 Apr 2016

%\cite{Mei:2002ip}
\bibitem{Mei:2002ip}
  Z.~H.~Mei and X.~Q.~Luo,
  %``Exotic mesons from quantum chromodynamics with improved gluon and quark actions on the anisotropic lattice,''
  Int.\ J.\ Mod.\ Phys.\ A {\bf 18}, 5713 (2003)
%  doi:10.1142/S0217751X03017038
  [hep-lat/0206012].
  %%CITATION = doi:10.1142/S0217751X03017038;%%
  %60 citations counted in INSPIRE as of 29 Apr 2016

%\cite{Hedditch:2005zf}
\bibitem{Hedditch:2005zf}
  J.~N.~Hedditch, W.~Kamleh, B.~G.~Lasscock, D.~B.~Leinweber, A.~G.~Williams and J.~M.~Zanotti,
  %``1-+ exotic meson at light quark masses,''
  Phys.\ Rev.\ D {\bf 72}, 114507 (2005)
%  doi:10.1103/PhysRevD.72.114507
  [hep-lat/0509106].
  %%CITATION = doi:10.1103/PhysRevD.72.114507;%%
  %43 citations counted in INSPIRE as of 29 Apr 2016

%\cite{Dudek:2010wm}
\bibitem{Dudek:2010wm}
  J.~J.~Dudek, R.~G.~Edwards, M.~J.~Peardon, D.~G.~Richards and C.~E.~Thomas,
  %``Toward the excited meson spectrum of dynamical QCD,''
  Phys.\ Rev.\ D {\bf 82}, 034508 (2010)
%  doi:10.1103/PhysRevD.82.034508
  [arXiv:1004.4930 [hep-ph]].
  %%CITATION = doi:10.1103/PhysRevD.82.034508;%%
  %151 citations counted in INSPIRE as of 29 Apr 2016

%\cite{McNeile:2006bz}
\bibitem{McNeile:2006bz}
  C.~McNeile {\it et al.} [UKQCD Collaboration],
  %``Decay width of light quark hybrid meson from the lattice,''
  Phys.\ Rev.\ D {\bf 73}, 074506 (2006)
%  doi:10.1103/PhysRevD.73.074506
  [hep-lat/0603007].
  %%CITATION = doi:10.1103/PhysRevD.73.074506;%%
  %101 citations counted in INSPIRE as of 29 Apr 2016

%\cite{Isgur:1983wj}
\bibitem{Isgur:1983wj}
  N.~Isgur and J.~E.~Paton,
  %``A Flux Tube Model for Hadrons,''
  Phys.\ Lett.\ B {\bf 124}, 247 (1983).
%  doi:10.1016/0370-2693(83)91445-4
  %%CITATION = doi:10.1016/0370-2693(83)91445-4;%%
  %187 citations counted in INSPIRE as of 29 Apr 2016

%\cite{Isgur:1984bm}
\bibitem{Isgur:1984bm}
  N.~Isgur and J.~E.~Paton,
  %``A Flux Tube Model for Hadrons in QCD,''
  Phys.\ Rev.\ D {\bf 31}, 2910 (1985).
%  doi:10.1103/PhysRevD.31.2910
  %%CITATION = doi:10.1103/PhysRevD.31.2910;%%
  %668 citations counted in INSPIRE as of 29 Apr 2016

%\cite{LeYaouanc:1984gh}
\bibitem{LeYaouanc:1984gh}
  A.~Le Yaouanc, L.~Oliver, O.~Pene, J.~C.~Raynal and S.~Ono,
  %``$q \bar{q} g$ Hybrid Mesons in $\psi \to \gamma$ + Hadrons,''
  Z.\ Phys.\ C {\bf 28}, 309 (1985).
%  doi:10.1007/BF01575740
  %%CITATION = doi:10.1007/BF01575740;%%
  %94 citations counted in INSPIRE as of 29 Apr 2016

%\cite{Ishida:1991mx}
\bibitem{Ishida:1991mx}
  S.~Ishida, H.~Sawazaki, M.~Oda and K.~Yamada,
  %``Decay properties of hybrid mesons with a massive constituent gluon and search for their candidates,''
  Phys.\ Rev.\ D {\bf 47}, 179 (1993).
%  doi:10.1103/PhysRevD.47.179
  %%CITATION = doi:10.1103/PhysRevD.47.179;%%
  %30 citations counted in INSPIRE as of 29 Apr 2016

%\cite{Iddir:1988jd}
\bibitem{Iddir:1988jd}
  F.~Iddir, A.~Le Yaouanc, L.~Oliver, O.~Pene, J.~C.~Raynal and S.~Ono,
  %``$q \bar{q} g$ Hybrid and $q q \bar{q} \bar{q}$ Diquonium Interpretation of Gams 1-+ Resonance,''
  Phys.\ Lett.\ B {\bf 205}, 564 (1988).
%  doi:10.1016/0370-2693(88)90999-9
  %%CITATION = doi:10.1016/0370-2693(88)90999-9;%%
  %66 citations counted in INSPIRE as of 29 Apr 2016

%\cite{Shifman:1978bx}
\bibitem{Shifman:1978bx}
  M.~A.~Shifman, A.~I.~Vainshtein and V.~I.~Zakharov,
  %``QCD and Resonance Physics. Theoretical Foundations,''
  Nucl.\ Phys.\ B {\bf 147}, 385 (1979).
 % doi:10.1016/0550-3213(79)90022-1
  %%CITATION = doi:10.1016/0550-3213(79)90022-1;%%
  %4694 citations counted in INSPIRE as of 29 Apr 2016

\bibitem{key-6-1}I.~I.~Balitsky, D.~Diakonov and A.~V.~Yung,
%  ``Exotic Mesons With $J^{pc} = 1^{-+}$ From Qcd Sum Rules,''
  Phys.\ Lett.\ B {\bf 112}, 71 (1982).
  %%CITATION = PHLTA,B112,71;%%
  %88 citations counted in INSPIRE as of 03 Nov 2014

\bibitem{key-7-1}J.~Govaerts, F.~de Viron, D.~Gusbin and J.~Weyers,
 % ``Exotic Mesons From {QCD} Sum Rules,''
  Phys.\ Lett.\ B {\bf 128}, 262 (1983).
  %%CITATION = PHLTA,B128,262;%%
  %46 citations counted in INSPIRE as of 03 Nov 2014

\bibitem{key-8-1}I.~I.~Balitsky, D.~Diakonov and A.~V.~Yung,
 % ``Exotic Mesons With $J^{pc} = 1^{-+}$, Strange and Nonstrange,''
  Z.\ Phys.\ C {\bf 33}, 265 (1986).
  %%CITATION = ZEPYA,C33,265;%%
  %63 citations counted in INSPIRE as of 03 Nov 2014

\bibitem{key-8-2}J.~Govaerts, F.~de Viron, D.~Gusbin and J.~Weyers,
 % ``{QCD} Sum Rules and Hybrid Mesons,''
  Nucl.\ Phys.\ B {\bf 248}, 1 (1984).
  %%CITATION = NUPHA,B248,1;%%
  %81 citations counted in INSPIRE as of 03 Nov 2014

\bibitem{key-8-3}J.~I.~Latorre, P.~Pascual and S.~Narison,
 % ``Spectra and Hadronic Couplings of Light Hermaphrodite Mesons,''
  Z.\ Phys.\ C {\bf 34}, 347 (1987).
  %%CITATION = ZEPYA,C34,347;%%
  %94 citations counted in INSPIRE as of 03 Nov 2014

\bibitem{key-8-4}J.~Govaerts, L.~J.~Reinders, P.~Francken, X.~Gonze and J.~Weyers,
%  ``Coupled {QCD} Sum Rules for Hybrid Mesons,''
  Nucl.\ Phys.\ B {\bf 284}, 674 (1987).
  %%CITATION = NUPHA,B284,674;%%
  %42 citations counted in INSPIRE as of 03 Nov 2014

\bibitem{key-8-5}S.~Narison,
%  ``Gluonia, scalar and hybrid mesons in QCD,''
  Nucl.\ Phys.\ A {\bf 675}, 54C (2000)
  [hep-ph/9909470].
  %%CITATION = HEP-PH/9909470;%%
  %50 citations counted in INSPIRE as of 03 Nov 2014

\bibitem{key-8-6}K.~G.~Chetyrkin and S.~Narison,
%  ``Light hybrid mesons in QCD,''
  Phys.\ Lett.\ B {\bf 485}, 145 (2000)
  [hep-ph/0003151].
  %%CITATION = HEP-PH/0003151;%%
  %46 citations counted in INSPIRE as of 03 Nov 2014

\bibitem{key-9}H.~Y.~Jin and J.~G.~Korner,
%  ``Radiative corrections to the correlator of ($0^{++}$,$1^{-+}$) light hybrid currents,''
  Phys.\ Rev.\ D {\bf 64}, 074002 (2001)
  [hep-ph/0003202].
  %%CITATION = HEP-PH/0003202;%%
  %9 citations counted in INSPIRE as of 03 Nov 2014

\bibitem{key-9-1} H.~Y.~Jin, J.~G.~Korner and T.~G.~Steele,
%  ``Improved determination of the mass of the $1^{-+}$ light hybrid meson from QCD sum rules,''
  Phys.\ Rev.\ D {\bf 67}, 014025 (2003)
  [hep-ph/0211304].
  %%CITATION = HEP-PH/0211304;%%
  %22 citations counted in INSPIRE as of 03 Nov 2014

%\cite{Narison:2009vj}
\bibitem{Narison:2009vj}
  S.~Narison,
  %``1-+ light exotic mesons in QCD,''
  Phys.\ Lett.\ B {\bf 675}, 319 (2009)
%  doi:10.1016/j.physletb.2009.04.012
  [arXiv:0903.2266 [hep-ph]].
  %%CITATION = doi:10.1016/j.physletb.2009.04.012;%%
  %20 citations counted in INSPIRE as of 29 Apr 2016

\bibitem{zhangthesis}
 Zhu-feng Zhang, ``QCD Sum Rules, Instanton and New Hadrons (in Chinese)'', Doctoral thesis, Zhejiang University, Hangzhou, China (2008).

%\cite{Zhang:2013rya}
\bibitem{Zhang:2013rya}
  Z.~f.~Zhang, H.~y.~Jin and T.~G.~Steele,
  %``Revisiting $1^{-+}$ and $0^{++}$ light hybrids from Monte-Carlo based QCD sum rules,''
  Chin.\ Phys.\ Lett.\  {\bf 31}, 051201 (2014)
%  doi:10.1088/0256-307X/31/5/051201
  [arXiv:1312.5432 [hep-ph]].
  %%CITATION = doi:10.1088/0256-307X/31/5/051201;%%
  %2 citations counted in INSPIRE as of 29 Apr 2016

%\cite{Huang:2014hya}
\bibitem{Huang:2014hya}
  Z.~R.~Huang, H.~Y.~Jin and Z.~F.~Zhang,
  %``New predictions on the mass of the $1^{-+}$ light hybrid meson from QCD sum rules,''
  JHEP {\bf 1504}, 004 (2015)
%  doi:10.1007/JHEP04(2015)004
  [arXiv:1411.2224 [hep-ph]].
  %%CITATION = doi:10.1007/JHEP04(2015)004;%%

%\cite{Chen:2008qw}
\bibitem{Chen:2008qw}
  H.~X.~Chen, A.~Hosaka and S.~L.~Zhu,
  %``The I**G J**PC = 1- 1-+ Tetraquark States,''
  Phys.\ Rev.\ D {\bf 78}, 054017 (2008)
 % doi:10.1103/PhysRevD.78.054017
  [arXiv:0806.1998 [hep-ph]].
  %%CITATION = doi:10.1103/PhysRevD.78.054017;%%
  %20 citations counted in INSPIRE as of 29 Apr 2016

%\cite{Zhang:2004nb}
\bibitem{Zhang:2004nb}
  Z.~F.~Zhang and H.~Y.~Jin,
  %``The Zero mode effect in the 1-+ four quark states,''
  Phys.\ Rev.\ D {\bf 71}, 011502 (2005)
 % doi:10.1103/PhysRevD.71.011502
  [hep-ph/0412226].
  %%CITATION = doi:10.1103/PhysRevD.71.011502;%%
  %4 citations counted in INSPIRE as of 29 Apr 2016


%\cite{Burns:2006wz}
\bibitem{Burns:2006wz}
  T.~Burns and F.~E.~Close,
  %``Hybrid meson properties in Lattice QCD and Flux Tube Models,''
  Phys.\ Rev.\ D {\bf 74}, 034003 (2006)
 % doi:10.1103/PhysRevD.74.034003
  [hep-ph/0604161].
  %%CITATION = doi:10.1103/PhysRevD.74.034003;%%
  %25 citations counted in INSPIRE as of 29 Apr 2016

%\cite{Kokoski:1985is}
\bibitem{Kokoski:1985is}
  R.~Kokoski and N.~Isgur,
  %``Meson Decays by Flux Tube Breaking,''
  Phys.\ Rev.\ D {\bf 35}, 907 (1987).
%  doi:10.1103/PhysRevD.35.907
  %%CITATION = doi:10.1103/PhysRevD.35.907;%%
  %386 citations counted in INSPIRE as of 08 May 2016

%\cite{Isgur:1985vy}
\bibitem{Isgur:1985vy}
  N.~Isgur, R.~Kokoski and J.~Paton,
  %``Gluonic Excitations of Mesons: Why They Are Missing and Where to Find Them,''
  Phys.\ Rev.\ Lett.\  {\bf 54}, 869 (1985)
  [AIP Conf.\ Proc.\  {\bf 132}, 242 (1985)].
  %doi:10.1103/PhysRevLett.54.869, 10.1063/1.35357
  %%CITATION = doi:10.1103/PhysRevLett.54.869, 10.1063/1.35357;%%
  %285 citations counted in INSPIRE as of 08 May 2016



%\cite{Page:1998gz}
\bibitem{Page:1998gz}
  P.~R.~Page, E.~S.~Swanson and A.~P.~Szczepaniak,
  %``Hybrid meson decay phenomenology,''
  Phys.\ Rev.\ D {\bf 59}, 034016 (1999)
  doi:10.1103/PhysRevD.59.034016
  [hep-ph/9808346].
  %%CITATION = doi:10.1103/PhysRevD.59.034016;%%
  %115 citations counted in INSPIRE as of 08 May 2016

%\cite{Swanson:1997wy}
\bibitem{Swanson:1997wy}
  E.~S.~Swanson and A.~P.~Szczepaniak,
  %``Hybrid decays,''
  Phys.\ Rev.\ D {\bf 56}, 5692 (1997)
  %doi:10.1103/PhysRevD.56.5692
  [hep-ph/9704434].
  %%CITATION = doi:10.1103/PhysRevD.56.5692;%%
  %30 citations counted in INSPIRE as of 08 May 2016

%\cite{DeViron:1985xn}
\bibitem{DeViron:1985xn}
  F.~De Viron and J.~Govaerts,
  %``Some Decay Modes Of 1-+ Hybrid Mesons,''
  Phys.\ Rev.\ Lett.\  {\bf 53}, 2207 (1984).
  %doi:10.1103/PhysRevLett.53.2207
  %%CITATION = doi:10.1103/PhysRevLett.53.2207;%%
  %39 citations counted in INSPIRE as of 29 Apr 2016

%\cite{Chen:2010ic}
\bibitem{Chen:2010ic}
  H.~X.~Chen, Z.~X.~Cai, P.~Z.~Huang and S.~L.~Zhu,
  %``The Decay Properties of the 1^{-+} Hybrid State,''
  Phys.\ Rev.\ D {\bf 83}, 014006 (2011)
 % doi:10.1103/PhysRevD.83.014006
  [arXiv:1010.3974 [hep-ph]].
  %%CITATION = doi:10.1103/PhysRevD.83.014006;%%
  %5 citations counted in INSPIRE as of 29 Apr 2016

%\cite{Huang:2010dc}
\bibitem{Huang:2010dc}
  P.~Z.~Huang, H.~X.~Chen and S.~L.~Zhu,
  %``The Strong Decay Patterns of the $1^{-+}$ Exotic Hybrid Mesons,''
  Phys.\ Rev.\ D {\bf 83}, 014021 (2011)
%  doi:10.1103/PhysRevD.83.014021
  [arXiv:1010.2293 [hep-ph]].
  %%CITATION = doi:10.1103/PhysRevD.83.014021;%%
  %5 citations counted in INSPIRE as of 29 Apr 2016

%\cite{Balitsky:1989ry}
\bibitem{Balitsky:1989ry}
  I.~I.~Balitsky, V.~M.~Braun and A.~V.~Kolesnichenko,
  %``Radiative Decay Sigma+ ---> p gamma in Quantum Chromodynamics,''
  Nucl.\ Phys.\ B {\bf 312}, 509 (1989).
 % doi:10.1016/0550-3213(89)90570-1
  %%CITATION = doi:10.1016/0550-3213(89)90570-1;%%
  %412 citations counted in INSPIRE as of 29 Apr 2016

%\cite{Braun:1988qv}
\bibitem{Braun:1988qv}
  V.~M.~Braun and I.~E.~Filyanov,
  %``QCD Sum Rules in Exclusive Kinematics and Pion Wave Function,''
  Z.\ Phys.\ C {\bf 44}, 157 (1989)
  [Sov.\ J.\ Nucl.\ Phys.\  {\bf 50}, 511 (1989)]
  [Yad.\ Fiz.\  {\bf 50}, 818 (1989)].
 % doi:10.1007/BF01548594
  %%CITATION = doi:10.1007/BF01548594;%%
  %363 citations counted in INSPIRE as of 29 Apr 2016

%\cite{Chernyak:1990ag}
\bibitem{Chernyak:1990ag}
  V.~L.~Chernyak and I.~R.~Zhitnitsky,
  %``B meson exclusive decays into baryons,''
  Nucl.\ Phys.\ B {\bf 345}, 137 (1990).
 % doi:10.1016/0550-3213(90)90612-H
  %%CITATION = doi:10.1016/0550-3213(90)90612-H;%%
  %376 citations counted in INSPIRE as of 29 Apr 2016

%\cite{Belyaev:1994zk}
\bibitem{Belyaev:1994zk}
  V.~M.~Belyaev, V.~M.~Braun, A.~Khodjamirian and R.~Ruckl,
  %``D* D pi and B* B pi couplings in QCD,''
  Phys.\ Rev.\ D {\bf 51}, 6177 (1995)
  %doi:10.1103/PhysRevD.51.6177
  [hep-ph/9410280].
  %%CITATION = doi:10.1103/PhysRevD.51.6177;%%
  %413 citations counted in INSPIRE as of 09 May 2016

%\cite{Leupold:1997dg}
\bibitem{Leupold:1997dg}
  S.~Leupold, W.~Peters and U.~Mosel,
  %``What QCD sum rules tell about the rho meson,''
  Nucl.\ Phys.\ A {\bf 628}, 311 (1998)
  doi:10.1016/S0375-9474(97)00634-9
  [nucl-th/9708016].
  %%CITATION = doi:10.1016/S0375-9474(97)00634-9;%%
  %142 citations counted in INSPIRE as of 29 Aug 2016
 %\cite{AliKhan:2001xoi}

\bibitem{AliKhan:2001xoi}
  A.~Ali Khan {\it et al.} [CP-PACS Collaboration],
  %``Light hadron spectroscopy with two flavors of dynamical quarks on the lattice,''
  Phys.\ Rev.\ D {\bf 65}, 054505 (2002)
  Erratum: [Phys.\ Rev.\ D {\bf 67}, 059901 (2003)]
  doi:10.1103/PhysRevD.65.054505, 10.1103/PhysRevD.67.059901
  [hep-lat/0105015].
  %%CITATION = doi:10.1103/PhysRevD.65.054505, 10.1103/PhysRevD.67.059901;%%
  %327 citations counted in INSPIRE as of 30 Aug 2016
%\cite{Close:1994hc}

%\cite{Jansen:2009hr}
\bibitem{Jansen:2009hr}
  K.~Jansen {\it et al.} [ETM Collaboration],
  %``Meson masses and decay constants from unquenched lattice QCD,''
  Phys.\ Rev.\ D {\bf 80}, 054510 (2009)
  %doi:10.1103/PhysRevD.80.054510
  [arXiv:0906.4720 [hep-lat]].
  %%CITATION = doi:10.1103/PhysRevD.80.054510;%%
  %36 citations counted in INSPIRE as of 21 May 2016

%\cite{Bakulev:1999gf}
\bibitem{Bakulev:1999gf}
  A.~P.~Bakulev and S.~V.~Mikhailov,
  %``QCD vacuum tensor susceptibility and properties of transversely polarized mesons,''
  Eur.\ Phys.\ J.\ C {\bf 17}, 129 (2000)
 % doi:10.1007/s100520000466
  [hep-ph/9908287].
  %%CITATION = doi:10.1007/s100520000466;%%
  %26 citations counted in INSPIRE as of 10 May 2016

\bibitem{reinder}
J. Govaerts , L J Reinders , F. Deviron and J. Weyers , Nucl. Phys. {\bf B283}, 706 (1987).

\bibitem{Close:1994hc}
  F.~E.~Close and P.~R.~Page,
  %``The Production and decay of hybrid mesons by flux tube breaking,''
  Nucl.\ Phys.\ B {\bf 443}, 233 (1995)
  %doi:10.1016/0550-3213(95)00085-7
  [hep-ph/9411301].
  %%CITATION = doi:10.1016/0550-3213(95)00085-7;%%
  %228 citations counted in INSPIRE as of 20 May 2016

%\cite{Ball:1998tj}
\bibitem{Ball:1998tj}
  P.~Ball,
  %``B ---> pi and B ---> K transitions from QCD sum rules on the light cone,''
  JHEP {\bf 9809}, 005 (1998)
 % doi:10.1088/1126-6708/1998/09/005
  [hep-ph/9802394].
  %%CITATION = doi:10.1088/1126-6708/1998/09/005;%%
  %319 citations counted in INSPIRE as of 29 Apr 2016

%\cite{Jansen:2009yh}
\bibitem{Jansen:2009yh}
  K.~Jansen {\it et al.} [ETM Collaboration],
  %``A Lattice QCD calculation of the transverse decay constant of the b(1)(1235) meson,''
  Phys.\ Lett.\ B {\bf 690}, 491 (2010)
  %doi:10.1016/j.physletb.2010.05.074
  [arXiv:0910.5883 [hep-lat]].
  %%CITATION = doi:10.1016/j.physletb.2010.05.074;%%
  %4 citations counted in INSPIRE as of 10 May 2016

%\cite{Reinders:1984sr}
\bibitem{Reinders:1984sr}
  L.~J.~Reinders, H.~Rubinstein and S.~Yazaki,
  %``Hadron Properties from QCD Sum Rules,''
  Phys.\ Rept.\  {\bf 127}, 1 (1985).
 % doi:10.1016/0370-1573(85)90065-1
  %%CITATION = doi:10.1016/0370-1573(85)90065-1;%%
  %1401 citations counted in INSPIRE as of 10 May 2016


%\cite{Narison:2002pw}
\bibitem{Narison:2002pw}
  S.~Narison,
  %``QCD as a theory of hadrons from partons to confinement,''
  Camb.\ Monogr.\ Part.\ Phys.\ Nucl.\ Phys.\ Cosmol.\  {\bf 17}, 1 (2001)
  [hep-ph/0205006].
  %%CITATION = HEP-PH/0205006;%%
  %194 citations counted in INSPIRE as of 05 Aug 2016


%\cite{Narison:2014wqa}
\bibitem{Narison:2014wqa}
  S.~Narison,
  %``Mini-review on QCD spectral sum rules,''
  Nucl.\ Part.\ Phys.\ Proc.\  {\bf 258-259}, 189 (2015)
  doi:10.1016/j.nuclphysbps.2015.01.041
  [arXiv:1409.8148 [hep-ph]].
  %%CITATION = doi:10.1016/j.nuclphysbps.2015.01.041;%%
  %5 citations counted in INSPIRE as of 09 Aug 2016
\end{thebibliography}
\end{document}